%% file: nota-heterotic-new.tex
\documentclass[12pt]{article}
\usepackage{cite}
\usepackage{amsmath, amsthm, amssymb,slashed,graphicx,epsfig}
\newcommand {\slsh} [1] {\not{\hbox{\kern-2pt${#1}$}}}

\newcommand{\gsim}{\lower.7ex\hbox{$\;\stackrel{\textstyle>}{\sim}\;$}}
\newcommand{\lsim}{\lower.7ex\hbox{$\;\stackrel{\textstyle<}{\sim}\;$}}
\newcommand {\beq} {\begin{equation}}
\newcommand {\eeq} {\end{equation}}
\newcommand {\beqn}{\begin{eqnarray}}
\newcommand {\eeqn} {\end{eqnarray}}
\newcommand{\bea}{\begin{eqnarray}}
\newcommand{\eea}{\end{eqnarray}}

\def\tr{{\rm Tr}\,}
\def\Tr{ \hbox{\rm Tr}\,}
\def\Trnc{ \hbox{\rm Tr} \, }

\def\tb{\widetilde{b}}
\def\tr{ \hbox{\rm Tr}}

\def\SU{{\rm SU}}

\def\te{\widetilde{e}}
\def\U{{\rm U}}

\def\tB{\widetilde{B}}

\def\tE{\widetilde{E}}
\def\te{\widetilde{e}}

\def\Re{\hbox {\rm Re}\,}
\def\Im{\hbox {\rm Im}\,}

\def\Z{\mathbb Z}
\def\1{\mathbbm{1}}
\def\N{{\cal N}}
\def\tq{{\widetilde q}}
\def\W{{\cal W}}
\def\tQ{{\widetilde Q}}
\def\dag{{}^{\dagger}}
\def\p{{}^{\,\prime}}
\def\a{\alpha}

\def\tnc{\widetilde{n}_c}
\def\tq{\widetilde{q}}
\def\snd{ \mbox{\tiny $ \N=2$ } }
\def\snu{ \mbox{\tiny $ \N=1$ } }

\def\CP{\rm CP}
\def\c{\rm c}
\def\f{\rm f}
\def\murg{\mu_{ \mbox{\tiny \rm RG }}}
\def\1{\mbox{\tiny (1) }}
\def\0{\mbox{\tiny (0) }}

\def\be{\begin{eqnarray}}
\def\ee{\end{eqnarray}}

\newcommand{\eqn}[1]{(\ref{#1})}

\def\nablaslash{\,\,{\raise.15ex\hbox{/}\mkern-12mu \nabla}}
\def\nablabarslash{\,\,{\raise.15ex\hbox{/}\mkern-12mu {\bar \nabla}}}
\def\Dslash{\,\,{\raise.15ex\hbox{/}\mkern-12mu D}}
\def\Dbarslash{\,\,{\raise.15ex\hbox{/}\mkern-12mu {\bar D}}}
\def\delslash{\,\,{\raise.15ex\hbox{/}\mkern-9mu \partial}}
\def\delbarslash{\,\,{\raise.15ex\hbox{/}\mkern-9mu {\bar\partial}}}
\def\pslash{\,\,{\raise.15ex\hbox{/}\mkern-9mu p}}
\def\calDslash{\,\,{\raise.15ex\hbox{/}\mkern-12mu {\cal D}}}

\def\lae{\mathrel{\mathop{\smash{\lower .5 ex \hbox{$\stackrel<\sim$}}}}}
\def\lae{\mathrel{\mathop{\smash{\lower .5 ex \hbox{$\stackrel>\sim$}}}}}

\def\beqn{\begin{eqnarray}}
\def\eeqn{\end{eqnarray}}

\def\ba{\beq\new\begin{array}{c}}
\def\ea{\end{array}\eeq}
\def\be{\ba}
\def\ee{\ea}

\newcommand{\ntwo}{${\mathcal N}=2\;$}
\newcommand{\ntwot}{${\mathcal N}= \left(2,2\right)\; $}
\newcommand{\ntwoo}{${\mathcal N}= \left(0,2\right)\; $}
\newcommand{\none}{${\mathcal N}=1\;$}
\newcommand{\vp}{\varphi}
\newcommand{\pt}{\partial}

\begin{document}

\begin{titlepage}

\begin{flushright}
FTPI-MINN-09/09; UMN-TH-2739/09\\3/19/09
\end{flushright}

\vskip 1.4in
\begin{center}
{\bf\large{On the Problem of the Quantum Heterotic Vortex}}\vskip0cm
\vskip 0.5cm {Stefano Bolognesi} \vskip
0.05in {\small{ \textit{William I. Fine Theoretical Physics Institute, University of Minnesota, } \\ {\textit{116 Church St. S.E., Minneapolis, MN 55455, USA}}}}
\end{center}
\vskip 0.5in

\baselineskip 16pt
%
%\date{February, 2009}

\begin{abstract}
We address the problem of non-Abelian super-QCD, with a Fayet-Iliopoulos term, as seen from the vortex worldsheet perspective.
Together with the FI term $\xi$, also a mass $\mu$  for the adjoint superfield $\Phi$ enters in the game. 
This mass allows the interpolation  between  $\N=2$ and $\N=1$ super-QCD. 
While the phenomenology of the $\N=2$ case ($\mu=0$) is pretty much understood, 
much remains to be clarified for the finite-$\mu$ case. 
We distinguish, inside the parameter space spanned by the FI term and the mass $\mu$, four different corners where some quantitative statements can be made. 
These are the regions where the strong dynamics can, in some approximation, be quantitatively analyzed.
We focus in particular on two questions: 1) Is the quantum vortex BPS or non-BPS? 2) What is the phase of the internal non-Abelian  moduli? 
We find that the answer to these questions  strongly depends upon the choice of the linear term in the superpotential. 
We also try to explain what happens in the most unexplored, and controversial, region of parameters, that of the quantum heterotic vortex, where   $\Lambda \ll \sqrt{\xi} \ll \mu$.

\end{abstract}

%\vfill
%\begin{flushleft}
%February 2009
%\end{flushleft}

\end{titlepage}
\vfill\eject

\tableofcontents
 
\section{Introduction}\label{intro}

\input{intro}

\section{Preliminaries}\label{p}

\input{p}

\section{Corner (A): Perturbed \boldmath$\N=2$}\label{a}

\input{a}

\section{Corner (B): Perturbed \boldmath$\N=1$}\label{b}

\input{b}

%\newpage
\section{Corner (C): Heterotic Vortex Theory}\label{c}

\input{c}

\section{Corner (D): Quantum  Heterotic Vortex}\label{d}

\input{d}

\section{Conclusion}\label{conclusion}

\input{concl}

%\appendix
%\section{Zero Modes}
%\input{z}

\section*{Acknowledgments}
I wish to thank A.~Yung and M.~Shifman for many useful discussions about the subject, and on the heterotic vortex problem in particular. 
I also want to thank A.~Vainshtein for stimulating conversations. I thank again M.S. for comments on the manuscript.
This work is supported by DOE grant  DE-FG02-94ER40823.

\end{document}

%% file: intro.tex
Since the first appearance of the non-Abelian vortex \cite{Hanany:2003hp, Auzzi:2003fs}, two major lines of research have been pursued. 
One, which is also the main focus of the present paper, is to understand, and make use of, the relationship between the four-dimensional gauge dynamics, and the wordsheet dynamics, of the zero modes confined to live on the vortex. 
A second line, which will be the main focus of   future work \cite{futuro}, is to understand the dynamics of non-Abelian magnetic monopoles in relation to that of the non-Abelian string.

Non-Abelian vortices  provide a way to map a four-dimensional non-Abelian gauge theory onto a  two-dimensional sigma-model.
The non-Abelian gauge theory has $n_c$ colors and $n_f$ flavors, and through the introduction of an appropriate Higgs breaking term, generally a Fayet-Iliopoulos (FI) term, lives in a so-called color-flavor locked vacuum, or root of the baryonic branch. 
Non-Abelian gauge theories in the color-flavor locked phase  have, in general, stable vortices, like the ordinary Abrikosov-Nielsen-Olesen flux tube, but
with the addition of a number of orientation internal modes, parametrized by the complex projective space $\CP(n_c-1)$. 
The low-energy dynamics of an infinite-length vortex is described by some variant of
the $\CP(n_c-1)$ sigma-model living on the $d=1+1$ dimensional worldsheet.
Additional modes, in particular the fermionic ones, live in line-bundles over $\CP(n_c-1)$.
The main focus of the paper will be on the case of the number of colors equal to the number of flavors: $n_c =n_f =n$.

In general, the low-energy theory on the
string worldsheet is split into two disconnected parts:
a free theory for (super)translational moduli and a nontrivial part, a theory of
interacting (super)orientational moduli, the $\CP(n-1)$ sigma-model.
The latter theory is in general completely fixed if there are unbroken supersymmetries.

The situation is by now quite clear for $\N=2$ SQCD \cite{Dorey:1998yh,Dorey:1999zk,Shifman:2004dr,Hanany:2004ea}. 
Due to holomorphic properties, quantities such as the mass of the BPS particles, do not depend upon the FI term.
This means that these invariant quantities can be computed in two different ways. 
One through the Seiberg-Witten solution of the original four-dimensional theory, and the other through the solution of the $\N=(2,2)$ that lives on the string worldsheet.

It was suggested in \cite{Shifman:2005st} the possibility of lowering the supersymmetries of the four-dimensional theory down to $\N=1$, but still keeping the BPS property of the string.
That can happen, classically, if we introduce a superpotential for the adjoint chiral superfield $W(\Phi)$, and we fine-tune the parameters so that one root of $W\p$ coincides with the mass of the quarks.
The theory on the vortex worldsheet preserves half of the supercharges of the four-dimensional theory, and that means that it is $\N=(0,2)$ or $\N=(2,0)$ depending on the vortex orientation. From here comes the name heterotic.
The number of fermionic zero modes though, does not change  because it is fixed by an index theorem. 
That means that these extra fermions remain  in the low-energy theory and must be described by a fermionic $\N=(0,2)$ multiplet.
In the work \cite{Edalati:2007vk}, it was shown how to construct the $\N=(0,2)$ theory living on the vortex worldsheet. 
The four-dimensional superpotential essentially enters as a worldsheet superpotential for the fermionic $\N=(0,2)$ superfield.
It is important here that there is a mixing between translational and orientational zero modes; otherwise, it would not be possible to break the $\N=(2,2)$ supersymmetries on the worldsheet \cite{Shifman:2005st}.
At least with a small value of the superpotential, the deformation of the $1+1$ theory can be simply obtained by the superpotential of the original four-dimensional theory with an appropriate coefficient that can be computed by a four-dimensional zero-mode overlap \cite{Edalati:2007vk, Shifman:2008wv}.

This paper is devoted to the study of the heterotic vortex, and its quantum-related problems. 
The analysis of the problem was initiated in \cite{quantumvortexstring,Tong:2008qd}, but many issues remain to be explored. 
The holomorphic properties discussed above for the $\N=2$ theory are now lost. 
We thus do not expect to find exact analytical agreement between quantities computed in various regions of the parameters' space. 
But we nevertheless expect to find qualitative agreement in various physical interesting questions.

We shall concentrate on the most studied theory, $\U(n)$ $\N=2$ gauge theory with the number of flavors equal to the number of colors. 
There are two possible deformations of the theory we shall be interested in. One is to give a mass to the adjoint fields, a superpotential term like
\beq
\label{supsenzalin}
\int d^2 \theta \ \sqrt{2}  \mu \ \Tr \frac{\Phi^2}{2} + {\rm h.c.} \ .
\eeq
The other deformation is a Fayet-Iliopoulos term
\beq
\label{fi}
- 2 \xi \int d^2\theta d^2 \bar{\theta} \ \tr \, V \ .
\eeq
All we shall do in this paper is  play with these two parameters, $\mu$ and $\xi$, and study the various dynamics that appear.

\begin{figure}[h!t]
\epsfxsize=8cm
\centerline{\epsfbox{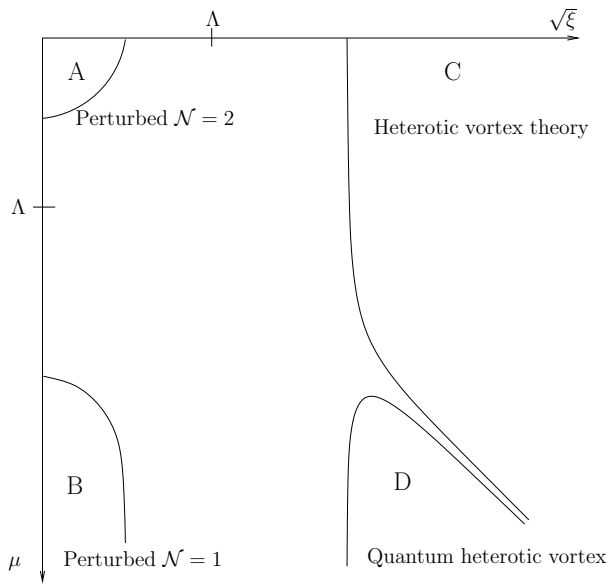}}
\caption{{\footnotesize We have two parameters to play with: the mass $\mu$ and the FI term $\sqrt{\xi}$. The four corners are the ones where some quantitative analysis can be carried out.}}
\label{box}
\end{figure}
In   Figure  \ref{box}, we present the square of the deformation parameters and, in particular, the four corners where the theory can be most easily  be analyzed. 
The goal of the paper is to analyze various phenomenological regimes of the strings,  at four different corners, try  to make sense of the various results.
Since the difficulties we shall deal with are those of strong-coupling regime, the only possible regions where we can safely analyze the theory  are in the four corners.
Let us now explain them briefly.
\begin{description}
\item[{\bf (A)}] This is the region where the theory is $\N=2$ essentially undeformed,  until energies well below the dynamical scale:
\beq
\sqrt{\xi}\ , \mu \ll \Lambda \ .
\eeq
Quantum effects can be treated using the Seiberg-Witten (SW) solution. 
Perturbations are then added to the thus-obtained low-energy effective Lagrangian.
We shall refer to this region as the perturbed $\N=2$, or perturbed SW  corner.
\item[{\bf (B)}]
Starting from the previous corner, we increase the mass $\mu$, until it becomes much greater than the dynamical scale: 
\beq
\sqrt{\xi} \ll \Lambda \ll \mu \ .
\eeq
We can then use the known results about strong dynamics of $\N=1$ SQCD.
At energies much smaller than the dynamical scales, the degrees of freedom are the gauge singlet mesons $\tQ Q$ and baryons $B=\epsilon Q \dots Q$, $\tB=\epsilon\tQ \dots \tQ$.
As long as we remain in the small FI limit, everything is weakly coupled, and we can introduce it as a perturbation to this low-energy effective theory. 
\item[{\bf (C)}]
As the FI term becomes larger, we eventually end up in the corner defined by:
\beq
\mu, \Lambda \ll \sqrt{\xi} \ .
\eeq
The $3+1$ dynamics is Higgsed at high energy, much above the dynamical scale. 
Dynamics survive  only inside the $1+1$ worldsheet of the non-Abelian vortex. 
The $\mu$ deformation is then added at low-energy.
This is the heterotic vortex string corner.
\item[{\bf (D)}]
The last corner is the one where both deformations are much grater than the dynamical scale;
\beq
\Lambda \ll \sqrt{\xi} \ll \mu \ .
\eeq
This is the most difficult region to analyze, and  less explored. 
Although the gauge $3+1$ dynamics is Higgsed at the weak coupling, not all particles in the bulk acquire mass. 
Some of them, corresponding to the mesonic moduli, have mass well below $\Lambda_{\snu}$.  
The worldsheet dynamics is generically entangled with these $3+1$ dimensional effects.
\end{description}

We shall first give a reasonable discussion of corners (A), (B), and (C). We shall also see that the various results agree, at least qualitatively.
Since the various corners are separated by regions where their respective treatments and approximations fail, we certainly cannot expect   quantitative  agreement between the various results. But at least we expect, and we shall find, a coherent explanation with respect to these two basic questions:
\begin{enumerate}
\item Is the vortex BPS or is the supersymmetry  broken?
\item What is the phase of  vortex  ground state? 
\end{enumerate}

Another important aspect,  is that the answers to the previous two questions  strongly depend  upon the choice of the linear term in the superpotential
\beq
\int d^2 \theta\ \sqrt{2} \mu \ \Tr \left(\frac{\Phi^2}{2} - a \Phi \right)+ {\rm h.c.} \ .
\label{sup}
\eeq
For the generic value of the coefficient $a$, the vortex is non-supersymmetric and the phase on its worldsheet is confining. For the particular case $a=0$, we have $n$ degenerate ground states,  with non-zero energy. Kinks correspond to the quanta of the orientational moduli,  and thus we lose confinement, but we still break supersymmetry.
For a particular choice, 
\beq
a \sim \Lambda e^{\frac{i \, 2\pi k}{n}} \ ,
\eeq
with $k= 1, \dots n$, one of the $n$ strings reaches zero energy. We thus have a BPS string with confinement. This superpotential is very peculiar because with this we {\it have} a supersymmetric heterotic vortex. We shall confirm this in all the three corners (A), (B), and (C).
In particular, we shall find enhancement of supersymmetry in the region (C).

As $\mu$ goes to infinity, some of the fermionic zero modes become broader and broader and finally non-normalizable in the $\N=1$ limit. The reason is the appearance of the additional massless particle, or equivalently, of the Higgs branch. 
It thus becomes tricky to disentangle the $3+1$ dynamics from the $1+1$ one. 
We cannot say, as we are used to in the $\N=2$ case, that at energy scales lower than the FI scale, all the $3+1$ excitations are gaped and the only non-trivial dynamics live on the string worldsheet. 
The $1+1$ theory becomes reliable at a scale much lower than the FI one. 
In particular for sufficiently large $\mu$, this scale becomes smaller than the dynamical scale. As $\mu \to \infty$, no $1+1$ approach can be said to be reliable.

The paper is organized as follows.
In Section \ref{p}, we present some preliminary material about the non-Abelian vortex-string. This is a review part;  we decided to collect the needed material in a mini-review, in order to facilitate the rest of the exposition.\footnote{See \cite{Tong:2005un,Shifman:2007ce} for more extensive reviews.}
In Sections \ref{a}, \ref{b}, \ref{c}, and \ref{d}, we discuss, respectively, the various corners in the parameters space (A), (B), (C), and  (D).
In the final Section \ref{conclusion}, we  summarize and conclude by discussing some of the future directions.

%% file: p.tex
We now provide some basic information and results about the non-Abelian theory and its non-Abelian vortex.
This Section is meant to be a quick review of the most important results we shall need in the bulk  of the paper.
We shall discuss the basic example, the $\U(n_c)$ $\N=2$ super-QCD with $n_f \geq n_c$ fundamental flavors.
We then consider the deformation of this theory with the FI term, and the superpotential for the adjoint field.
The non-Abelian string arises in the color-flavor locked vacuum, also-called the root of the baryonic branch.
We  focus in particular on the $n_c =n_f =n$ case, and introduce the non-Abelian $\CP(n-1)$ moduli that live on the vortex worldsheet.
We discuss the basic techniques to detect and study the bosonic and fermionic zero modes of the string. 
We in particular focus on the case of coinciding quark masses and the root of the derivative of the superpotential.
This is the case when the vortex preserves half of the supercharges, and the effective worldsheet theory is the $\N=(0,2)$ heterotic vortex theory.
On the way, we  refer  to the literature, where more detailed descriptions of each topic can be found.

\begin{center}
 * * *
\end{center}
The non-Abelian $\U (n_c)$ $\N=2$ theory  has gauge vector multiplet $W_{\alpha},\Phi$ and $n_f$ fundamental hypermultiplets $Q,\tQ^\dagger$. The physical  fields are, respectively,
\beq
\begin{array}{ccc}
 &A_{\mu}&\\
\lambda&&\psi\\
&\phi&
\end{array} \qquad \qquad 
\begin{array}{ccc}
 &\psi_q&\\
q&&\tq^{\,\dagger}\\
&\psi^{\dagger}_{\tq}&
\end{array} \ .
\eeq
The Lagrangian in the $\N=1$ superfield formulation is as follows:
\bea
\label{sqcdN=2}
{\cal L}&=& \int d^2\theta d^2\bar{\theta} \,  \frac{2}{g^2} \Trnc (\Phi^\dagger e^V \Phi e^{-V})+ \sum_{i=1}^{n_f}\, ( Q_i^\dagger e^{V}Q^i+\widetilde{Q}_i e^{-V} {\widetilde{Q}}^{\dagger i} ) +\nonumber\\
&&+\Im \int d^2\theta \,  \frac{ \tau}{4 \pi } \Trnc (W^{\alpha}W_{\alpha})+ \left[ \int d^2\theta \,  \W(\Phi,Q,\tQ)  + {\rm h.c.}  \right]  \ ,
\eea
where the coupling is
\beq
\tau =  \frac{4 \pi i}{ g^2} + \frac{\theta}{2\pi}\ ,
\eeq
the superpotential is
\beq
\W(\Phi,Q,\tQ)=\sum_{i=1}^{n_f} \, \sqrt{2}(\widetilde{Q}_i\Phi Q^i - m\widetilde{Q}_i Q^i ) \ ,
\eeq
and $m$ are the masses for the flavors with the index $i=1,\dots,n_f$.
We shall  consider only the case of degenerate masses. The parameter $m$ can  be absorbed in a shift of the coordinate $\Phi$, but we shall keep it explicit.

We can break half supersymmetries, $\N=2$ down to $\N=1$ , by adding a superpotential $W(\Phi)$ for the adjoint field  $\Phi$.  The total superpotential becomes
\beq
\W(\Phi,Q,\tQ)=\sum_{i=1}^{n_f} \, \sqrt{2}(\widetilde{Q}_i\Phi Q^i - m\widetilde{Q}_i Q^i )  +  \sqrt{2} \Trnc W(\Phi) \ ,
\eeq
where $W(z)$ is a generic holomorphic function. We shall be interested only in the quadratic superpotential
\beq
W(z)=\mu\Tr\left(\frac{\Phi^2}{2} - a \Phi\right)\ .
\eeq
The linear term, with coefficient $a$, shall play an important role in what follows.

The running of the coupling constant is given by
\beq
\Lambda^{\gamma n_c -n_f} = \murg^{\gamma n_c -n_f}  e^{-2\pi i \, \tau(\murg)} \ ,
\eeq
where $\Lambda$ the dynamical scale,  $\murg$ the renormalization group scale, and  $\gamma = 2$ or $3$ for, respectively, $\N=2$ and $\N=1$ .
The matching of the dynamical scales between the $\N=2$ and  $\N=1$ theories is given by
\beq
\label{relationofscales}
\Lambda_{\snu}^{3n_c-n_f} = \mu^{n_c} \Lambda_{\snd}^{2n_c-n_f}\ .
\eeq
This can be inferred by  running of the coupling constants or equivalently by the $\U(1)_R$ anomaly charges. 
For the case we shall be most interested in, this is $\Lambda_{\snu}^{2} = \mu \Lambda_{\snd}$.
\begin{center}
 * * *
\end{center}
We now add the Fayet-Iliopoulos term (\ref{fi}). This term lifts completely the Coulomb branch of the moduli space, leaving only the vacuum where $\phi= 0$ and the Higgs branch attached to it.
In the case of coinciding root of $W\p$ and the quarks masses, $a=m$,
we have:
\bea
\langle q \rangle &=& \left(\begin{array}{ccccc}
\sqrt{{2\xi}}&&&0&\\
&\ddots&&&\ddots\\
&&\sqrt{{2\xi}}&&\\
\end{array}\right) \ , \nonumber \\
\langle \tq \rangle &=& 0 \ .\label{higgsbreaking}
\eea
The breaking of the global symmetries of the Lagrangian is
\beq
\label{ressimm}
\SU(n_c) \times \SU(n_f) \rightarrow \SU(n_c)_{\rm c+f} \times \U(n_f-n_c) \ .
\eeq
The theory lies in the
color-flavor locked phase, with the vacuum expectation value
preserved by a simultaneous gauge and flavor rotation; from here comes the subscript c+f. 
%\beq
%\U(n)_{\c} \times \SU(n)_{\c} \rightarrow \SU(n)_{\c+\f}\ %.\label{breaking}
%\eeq

To get the masses of the scalar bosons, we expand the potential 
near the vacuum,  and diagonalize the
corresponding mass matrix \cite{Shifman:2005st}. The scalar $q$ becomes a partner of the  massive gauge bosons through the super-Higgs mechanism, and has mass
\beq
m_{A_{\mu}} = g \sqrt{2 \xi} \ .
\eeq
Other  scalar  fields $\tilde{q}$, $\phi$ acquire mass
too.
%\beq
%m^{+}=g\sqrt{\xi\lambda^{+}} \ , \qquad
%m^{-}=g\sqrt{\xi\lambda^{-}}\ 
%\label{u1m2}
%\eeq
%where $\lambda^{\pm}$ are two roots of the quadratic equation
%\beq
%\lambda^2-\lambda\left( 2+ \frac{g^2\mu^2}{\xi} \right) +1=0\,,
%\label{queq}
%\eeq 
%Here we introduced two \ntwo supersymmetry breaking
%parameter
%\beq
%\omega=\frac{g\mu}{\sqrt{\xi}} \ .
%\eeq 
%Note that all states come either as  singlets
%or triplets of unbroken SU(n-1)$_{\rm c+f}$.
When the $\N=2$ supersymmetry breaking vanishes, the
masses   coincide with the  gauge
boson mass  and the corresponding states form the bosonic part of the \ntwo
long massive vector supermultiplet.  
With non-zero
$\mu$, this supermultiplet splits into a massive vector multiplet and two chiral  heavy and light multiplets.

In the limit
of large \ntwo supersymmetry breaking $\mu \gg g\sqrt{\xi}$,  these light and heavy masses become
\beq
\label{masseshl}
m_{\rm h} = \sqrt{2} \, g \mu \ , \qquad  \qquad m_{\rm l} = 2 \sqrt{2} \, \frac{\xi}{\mu} \ .
\eeq
The heavy one, $m_{\rm h}$,  is the mass of the heavy adjoint
scalar $\phi$. 
Integrating out the adjoint field, we have $\Phi=Q\tQ/\mu + a$, and the effective superpotential is
\beq
\W_{\rm eff} = - \sqrt{2} \tr \left(  \frac{1}{2\mu} Q \tQ \, Q \tQ + (m-a) Q\tQ +\frac{\mu a^2}{2}
 \right) \ .
\label{effsup}
\eeq
In  the limit of infinite $\mu$,  the $m_{\rm l}$ masses tend to zero.
This fact reflects the enhancement of the Higgs branch in \none SQCD.

%In the $\mu \to \infty$ limit all $F$ terms disappear and we are left only with $D$ terms.
%Clearly in this limit the Higgs branch becomes enhanced.
%In particular the $F_{\phi}$ term, that provides a mass for the mesonic field $\tQ Q$, becomes ineffective because the $\phi$ field, when integrated out, leaves only a $1/\mu$ remnant effect (\ref{effsup}).

%\texttt{explain the relation with the case without FI term.}

The same thing happens   when the FI term is absent \cite{Bolognesi:2008sw}. The only difference is that at the base of the Higgs branch $m_{\rm l}$ is always zero. The $\tq q$ direction is lifted, but by a sixth-order potential.

\begin{center}
 * * *
\end{center}
Now we study the non-Abelian BPS vortex at the root of the Higgs baryonic branch. We set $\phi=0$ and $\tq=0$, and ignore any interference of these fields for the moment. 
We can certainly do that in the case where the superpotential is absent.
The part of the Lagrangian we are interested in is
\beq
\label{lagvortice}
{\cal L} = -\frac{1}{2g^2} \Tr \, (F_{\mu\nu}F^{\mu\nu}) + (D_\mu q)^{\dagger}(D^\mu q) - \frac{g^2}{4} \Tr \,(qq\dag - 2 {\xi} \, {\bf 1}_{n})^2\ .
\eeq
The central $\U(1) \subset \U(n_c)$ does not survive the breaking (\ref{higgsbreaking}), and this provides sufficient topology
to ensure the presence of the vortex  in the theory.

We write the tension of the vortex using the Bogomolny trick
\beq
T=T_{\rm BPS}+ \int d^2x  \, \frac12 (D_k q+i\epsilon_{kl}D_l q)\dag (D_k+i\epsilon_{kl}D_l q)
+\frac12 \Tr\, (\frac 1g F_{kl}+\frac{g}{2}(qq\dag -2{\xi})\epsilon_{kl})^2 \ ,
\eeq
where the lower bound is a boundary term
\beq
T_{\rm BPS}=2{\xi} \oint d \vec x \cdot \Tr\, \vec A \ .
\eeq
This term saturates the tension when the non-Abelian Bogomolny equations
\beq
D_kq+i\epsilon_{kl}D_l q=0 \ ,\qquad \frac 1g F_{kl}+\frac{g}{2}(qq\dag -2{\xi})\epsilon_{kl}=0
\eeq
are satisfied.

To build the vortex configuration, we embed the ordinary $\U(1)$ vortex in this theory.  All such embeddings are $\U(n_c)$ rotations of
\bea
\label{vortexconf}
q&=&\left(\begin{array}{cccccc}
e^{i\theta}  q(r)  \sqrt{2{\xi}} &&&&0&\\
&\sqrt{2{\xi}}&&&&\ddots\\
&&\ddots&&&\\
&&&\sqrt{2{\xi}}&&\\
\end{array}\right)\ ,\\
A_k&=&\left(\begin{array}{cccc}
-\epsilon_{kl}\frac{\hat{r}_l}{r}f(r)&&&\\
&0&&\\
&&\ddots&\\
&&&0\\
\end{array}\right) \ ,\nonumber
\eea
where $q(r)$ and $f(r)$ are some profile functions that satisfy the boundary conditions $q(0)=f(0)=0$ and $q(\infty)=f(\infty)=1$.
The $n_c$ independent vortices constructed this way are degenerate with tension. These are the non-Abelian vortex equations.
Solutions to these equations have tension
\beq
 T= 4 \pi    \xi 
\eeq
Note that this is $1/n$th of the tension of the ANO vortex, $T_{\rm ANO} = 4 n \pi \xi$, which is
obtained by a simultaneous winding of all the diagonal components of $q$.
%\beq
%\pi_1\left( \frac{\SU(n-1) \times \U(1)}{\Z_{n-1}} \right) = \Z
%\eeq

The vortex solution (\ref{vortexconf}) classically breaks the residual global symmetry (\ref{ressimm}). This leads to the existence of a moduli space. When $n_f=n_c$, we can find other equivalent solutions by taking the (\ref{vortexconf}) and making the following transformation
\beq
q \rightarrow U_{\c} \ q \ U_{\f}^{-1}\ , \qquad A_{\mu} \rightarrow U_{\c}\  A_{\mu} \ U_{\c}^{-1} \ ,
\eeq
with a color-flavor locked rotation $U_{\c} = U_{\f}$ in order to preserve the asymptotic vacuum. 
The moduli space of solution is then given by the coset space
\beq
\CP(n-1)=\frac{ \SU(n)_{\rm c+f} }{ \U(1) \times \SU(n-1)} \ .
\eeq
We can express the vortex solution in a way that makes manifest these bosonic moduli, by going to the singular gauge:
\bea q^a_{\ i} &=& \left(\frac{ \vp^a \bar{\varphi}_i}{\beta} \right)\,
\sqrt{2 \xi} (q(r)-1) + \sqrt{2\xi}\delta^a_{\ i} \ , \nonumber \\ [3mm]
(A_k)^a_{\ b} &=&   \left( \frac{\vp^a \bar{\varphi}_b}{\beta}\right) \epsilon_{kl}\frac{\hat{r}_l}{r} (1- f(r) )  \ .
\label{solutions}
\eea
The $\vp^l \in {\mathbb C}^{n_c}$ defines the orientation of the
vortex in the gauge and flavor groups. 
We require
\beq \sum_{i=1}^{n} |\vp^l|^2 = \beta \ ,
\label{constraint}
\eeq
with $\beta$ a constant that will be fixed in order to have canonical normalization for the kinetic term of $\vp$. 
The solutions
\eqn{solutions} are invariant under the simultaneous rotation,
\beq \vp^l \rightarrow e^{i\alpha}\vp^l\ .\label{iden}\eeq
The $\vp^i$, subject to the constraint (\ref{constraint}) and
identification \eqn{iden}, provide homogeneous coordinates on
the projective space ${\CP}(n_c-1)$. 
The $\SU(n_c)$ symmetry of four dimensions
descends to the vortex string, with the $\vp^i$ transforming in
the fundamental representation. 
%This ensures that the ${\bf
%CP}^{N_c-1}$ is endowed with the symmetric Fubini-Study metric.
%The K\"ahler class of this space is $r$.

When $n_f>n_c$, other zero modes are present in our theory. In particular, among the classical solutions for $n_f>n_c$, there are semi-local vortices.
These solitons interpolate between Nielsen-Olesen-like vortices
and sigma-model lumps on the Higgs branch of the theory. 
\begin{center}
 * * *
\end{center}
In the case of no-coincidence, i.e. $W\p(m) \neq 0$, we cannot neglect the $\tq$ field.
In the limit of a small superpotential though, we can still perform the computation just replacing $\phi=0$. 
This is an approximation that becomes exact in the limit of small superpotentials \cite{Auzzi:2004yg}.
The scalar potential of the theory is  
\beq
V= g^2 \tr \, |q\tq  + W'(m)|^2 + \frac{g^2}{4} \Tr \, (qq\dag -\tq\dag \tq- 2 \xi )^2 \ ,
\eeq
This may be expressed in an $\SU(2)_R$ invariant form using the doublet $q^{\alpha}=(q,\tq\dag)$:
\beq
V=\frac{g^2}{2}\Tr_{n}  \Tr_2 \, ({q\dag}^{\alpha}q_{\beta}-\frac{1}{2}\delta^{\alpha}_{\beta}{q\dag}^{\gamma} q_{\gamma}-\zeta_a(\sigma_a)^\alpha_\beta)^2\ , \eeq
where $\zeta_a$ is the $\SU(2)_R$ triplet defined by
\beq
 - \zeta_1 +  i\zeta_2= W\p(m)\ ,\qquad \zeta_3=\xi\ .
\eeq
An $\SU(2)_R$ rotation brings the potential to a  form with a new FI term and no superpotential
\beq
V=\frac{g^2}{4}\Tr \, (qq\dag -\tq\dag \tq-2{\xi\p})^2\ ,
\label{rotated}
\eeq
where $\xi\p=\sqrt{ |W\p|^2+{\xi^2}}$.

We can now use the rotated potential (\ref{rotated}) and write a solution for the vortex. The tension will be
\beq
T=4\pi \sqrt{ |W\p|^2+{\xi^2}} \ . \label{newv}
\eeq
As explained in \cite{Auzzi:2004yg}, in the limit 
\beq
g^2\left|{{W\p}\p}^2 {W\p}^2\right| \ll \left(|{W\p}|^2+ \xi^2\right)^{3/2}\ ,
\label{condition}
\eeq
we can neglect the $\phi$ field and use the potential (\ref{rotated}) to construct an almost-BPS vortex with tension $4 \pi \xi \p$. For $W$ sufficiently small, the condition (\ref{condition}) is satisfied. 
The point is that non-BPS corrections vanish faster than the main BPS contribution.
We shall use this trick in Section \ref{a}, to compute the vortex tension in the SW regime.
\begin{center}
 * * *
\end{center}
Now we will  study  the effective low-energy theory on the string worldsheet.
For the basic Abrikosov-Nielsen-Olesen vortex, this consists just of the tranversal fluctuation of the vortex, when embedded in space-time. The zero mode is the one generated by the broken translations operators, acting on the vortex solution. The coefficient in front of the kinetic term, is just the tension of the vortex. In the low-energy limit, these fluctuations are just described by free scalar and fermionic fields on the worldsheet.

The most interesting feature of the non-Abelian vortex is the presence of orientational zero modes $\vp^i$, and these too we must be considered in an effective low-energy description of the soliton fluctuations. Since these are parametrized by a $\CP(n-1)$ space, we expect a version of the $\CP(n-1)$ model. What makes it interesting is that the dynamic  in the infrared is now non-trivial and, generally, strongly coupled.

Assume  that the orientational collective coordinates $\vp^i$
are   slow varying functions of the string worldsheet coordinates
$x_{0,3}$. 
Then $\vp^l$ become fields in a (1+1)-dimensional
sigma model on the worldsheet. 
Since   the vector   $\vp^l$ parametrizes the string zero modes,
there is no potential term in this sigma model. The effective action, in the so-called gauged formulation, is thus
\beq
S_{1+1}
 =
\int d^2 x \Big\{
 |\nabla_{k} \vp^{l}|^2  +  D (|\vp^{l}|^2 - \beta)
\Big\}\,,
\label{cp}
\eeq
where 
$
\nabla_k= \partial_k - i A_k 
$. The auxiliary gauge field $A_{\mu}$ is necessary in order to make the phase (\ref{iden}) unphysical.  Eliminating the $D$ auxiliary field leads to the constraint
(\ref{constraint}). 
As we said, we chose  to normalize $\vp^l$ in order to have a canonical kinetic term. In this way, the inverse of the coupling constant appears as the radius of the $\CP(n-1)$ manifold, $r^2 =\beta$.

% \beq
%\beta = \frac{2\pi }{g^2}
%\eeq

There are also other formulations of this model, in particular the so-called geometric formulation. 
The last has no auxiliary fields, and the interactions are explicitly given by the geometry of the sigma-model manifold. The gauge formulation is the only one that we shall use in this paper. It is particularly useful because the auxiliary fields become dynamical through quantum corrections, and, in particular, it can be solved exactly in the large-$n$ limit.

There is an important issue we want to stress. 
The zero mode analysis can allow us to compute the coefficient in front of the kinetic term. Symmetries allow us to complete the Lagrangian, and infer the structure of interactions.

The study of the vortex theory makes sense only if the four-dimensional theory is still at weak coupling when the Higgs breaking happens. The condition is
\beq
\label{betaatxi}
\beta_0 = \frac{2 \pi}{g(\sqrt{\xi})^2} = \frac{n}{4 \pi} \log \frac{\sqrt{\xi}}{\Lambda_{3+1}} + \dots \gg 1 \ .
\eeq
The coupling of the vortex theory is  fixed by the coupling of the four-dimensional theory, but computed at the scale of the vortex, that is, $\sqrt{\xi}$. 
A classical computation provides the relationship between the worldsheet coupling $\beta$ and the four-dimensional gauge coupling \cite{Auzzi:2003fs}.

What happens at energy scales lower than $\sqrt{\xi}$, depends on the details of the theory. 
Usually, {\it but not always}, the $3+1$-dimensional degrees of freedom are gaped and do not alter the vortex theory below the scale $\sqrt{\xi}$.
It is just a simple passage of the baton, the $3+1$ dynamics runs until the scale $\sqrt{\xi}$, and then the $1+1$ dynamics starts its running. The two dynamical scales are essentially the same
\beq
\Lambda_{\rm CP}  \sim \Lambda_{3+1} \ .
\eeq
There is an important exception to this, to be discussed in Section \ref{d}.
\begin{center}
 * * *
\end{center}
We now turn to a study of the fermionic zero modes in the vortex background.
We will pay particular
attention to the correlation between the  chirality of the
worldsheet and that of the four-dimensional fermions.
These issues become  very important in the study of the heterotic vortex \cite{Edalati:2007vk}.

The ${\cal N}=2$ vector multiplet in four dimensions contains the two
Weyl fermions, $\lambda$ and $\psi$.  Each hypermultiplet
also contains two Weyl fermions, $\psi_q$ and $\psi_{\tq}$. 
We need to consider the Dirac equations in the background of the  vortex:
\bea
-\frac{i}{g^2} \nablabarslash \lambda
-\frac{i\sqrt{2}}{g^2}[\bar{\psi},\phi]
+i\sqrt{2}q_i\bar{\psi}_{q \, i}-
i\sqrt{2}\bar{{\psi}}_{\tq \, i}\widetilde{q}_i&=&0 \nonumber \\
-\frac{i}{g^2} \nablabarslash \psi - \frac{i\sqrt{2}}{g^2}
[\phi,\bar{\lambda}]-\sqrt{2}\widetilde{q}^{\,\dagger}_i\bar{\psi}_{q \, i}-
\sqrt{2}\bar{{\psi}}_{\tq \, i} {q}_i^{\dagger} &=& \sqrt{2} W^{''}({\phi})  \bar{\psi}
\label{diracuno}
\eea
and 
\bea
-i\nablabarslash\psi_{q\,i}+i\sqrt{2}\bar{\lambda}q_i-\sqrt{2}\phi^\dagger\bar{{\psi}}_{\tq \, i}
-\sqrt{2}\bar{\psi}\widetilde{q}_i^{\,\dagger} &=&0 \nonumber \\[4mm] 
-i\nablabarslash
{\psi}_{\tq\, i}-i\sqrt{2}\widetilde{q}_i\bar{\lambda}-\sqrt{2}\bar{\psi}_{q\,i}
\phi^\dagger -\sqrt{2}q_i^{\dagger}\bar{\psi}&=& 0 \ .
\label{diracdue}
\eea
The superpotential $W(\phi)$ enters only in the Dirac equation for $\psi$, the superpartner of $\phi$.
We consider the vortex to be static, and oriented in the positive $x^3$ direction. 
We decompose the spinors as
$(\lambda_L,\lambda_R)$. The derivative is
\beq
\nablabarslash = (\bar{\sigma}^{\mu})^{\alpha \dot{\alpha}} \nabla_{\mu} = - \left(
\begin{array}{ll}
\nabla_0+\nabla_3=\nabla_R & \nabla_1 -i \nabla_2 =\nabla_z\\ 
\nabla_1+i\nabla_2=\nabla_{\bar{z}} & \nabla_0-\nabla_3 =  \nabla_R
\end{array}
\right)
\eeq
We call $\lambda_R$ the {\it right movers} since $\nabla_L$ acts on them, and $L$ the {\it left movers} because $\nabla_R$ acts instead.
Fermionic zero modes of the Dirac equations (\ref{diracuno}) and (\ref{diracdue})  must then be interpreted as massless fermions localized on the $1+1$ vortex effective action.

In the $\N=2$ limit, we have $W(\phi) = 0$, and $\phi=\tilde{q}_i=0$ for the vortex solution.
The equations decouple into two pairs: the first set of equations
are for $\lambda$ and $\psi_{q\,i}$, 
and the second set of equations are for $\psi$ and
${{\psi}}_{\tq\,i}$.
Each  pair of four-dimensional fermions gives a Fermi zero
mode on the vortex of a specific worldsheet chirality. 
In particular, zero modes of the first pair live  in the components $\lambda = (0,\lambda_R)$, $\psi_{q\,i}=(\psi_{q\,L\,i},0)$, and give the right-handed fermions on the vortex worldsheet.
All the zero modes of the second pair live in the components $\psi=(\psi_L,0)$, $\psi_{\tq\,i}=(0,\psi_{\tq\,R\,i})$ and give the left-handed fermions.
Each of these pairs of equations, as a consequence of supersymmetry, is the same as the equations for
bosonic zero modes, derived by linearizing
the vortex equations and imposing a gauge fixing constraint. 
%The
%relationship between the bosonic and fermionic  zero modes is
%given by
%$\lambda_R \leftrightarrow \delta A_{\bar{z}}$, %$\sqrt{2}\bar{\psi}_{q\,i\, R}\leftrightarrow -i\delta q_i^\dagger$, %$\psi_L \leftrightarrow \delta A_z$,
%$\sqrt{2}\bar{{\psi}}_{\tq\, i\, L}\leftrightarrow -\delta
%q_i $.
%This mapping between the zero mode profiles is a consequence of
%the preserved supersymmetry in the background of the vortex. 

The low-energy dynamics of the vortex string arise  by promoting
the collective coordinates $z$, $\zeta_{R,L}$ and  $\vp^l$,  $\xi_{R,L}^l$ to dynamical fields on the string worldsheet. 
The fact that the vortices
are BPS, preserving 1/2 of the ${\cal N}=2$ four-dimensional
supersymmetry, ensures that the resulting worldsheet dynamic  is
invariant under  ${\cal N}=(2,2)$ supersymmetry. 
The
various bosonic and fermionic collective coordinates are grouped into ${\cal N}=(2,2)$ chiral superfields $Z$ and $\Phi^l$:
% The translational mode
%$z$ and the two Goldstino modes $\zeta_{R,L}$ sit in an ${\cal
%N}=(2,2)$ chiral multiplet $Z$. 
\beq
\begin{array}{ccc}
 & z &\\
\zeta_L && \zeta_R \\
\end{array}\ , \qquad \qquad
\begin{array}{ccc}
 &\vp^l&\\
\xi^l_L && \xi^l_R\\
\end{array} \ .
\eeq 
The constraint $\bar{\vp}_l \vp^l = \beta$  together with the identification (\ref{iden})  are imposed on the worldsheet theory by introducing
an auxiliary ${\cal N}=(2,2)$ vector supermultiplet: 
\beq
\begin{array}{ccc}
 & \sigma &\\
\chi_L && \chi_R \\
& A_{\mu} &\\
\end{array}\ ,
\eeq
That also will impose a constraint on the fermion $\xi_{l (L,R)}$.

We now give the explicit formulation for the  $\N=(2,2)$ $\CP(n-1)$ sigma-model. We skip the super field formulation, which can be found in the given references, and just present the final result in terms of fields.
The bosonic part of the action takes the form
\beqn
S_{\CP \,{\rm bos}}
& =&
\int d^2 x \left\{
 |\nabla_{k} \vp^{l}|^2  - \frac1{4e^2}F^2_{kl} + \frac1{e^2}
|\pt_k\sigma|^2 + \frac1{2e^2}D^2
\right.
\nonumber\\[3mm]
 &&   - 2|\sigma|^2 |\vp^{l}|^2 + D (|\vp^{l}|^2-\beta)
\Big\}\,,
\label{cpg}
\eeqn
where 
$ F_{kl} = \partial_k A_l -
\partial_l A_k$.
 $\sigma$ is a complex scalar field that is necessarily part of the gauge supermultilpet containing $A_{\mu}$, and $D$ is the $D$-component of the gauge multiplet.
We also wrote explicitly a gauge kinetic term for the gauge supermultiplet. In the  limit $e^2\to\infty$,
the gauge field $A_k$  and its \ntwo bosonic superpartner $\sigma$ become
auxiliary (their kinetic terms vanish) and can be eliminated by virtue of the equations of motion. The classical action has no gauge kinetic term ($e^2 = \infty$). It will be generated, in general, by quantum corrections.

The fermionic part of the $\CP(n-1)$ model action 
has the form
\beqn
S_{\CP \,{\rm ferm}}
& =&
\int d^2 x \Big\{
 \bar{\xi}_{R\,l} \,i \nabla_{L} \xi^{l}_R
+ \bar{\xi}_{L\,l}\,i \nabla_{R} \xi^{l}_L
+\frac1{e^2}\,\bar{\chi}_{R}\,i \nabla_{L} \chi_R
+\frac1{e^2}\,\bar{\chi}_{L}\,i \nabla_{R} \chi_L  \Big.
\nonumber\\[3mm]
&&
\Big.
-\sqrt{2}\,\bar{\sigma}\,\bar{\xi}_{l\,R}\xi^l_L
-\sqrt{2}\,\bar{\vp}_l\,(\xi^l_L \chi_R -
\xi^l_R \chi_L) + {\rm h.c.}  
\Big\} \ ,
\label{cpgf}
\eeqn
where the fields $\xi^l_{L,R}$ are fermion superpartners of $n^l$ while $\chi_{L,R}$
belong to the gauge multiplet. In the limit $e^2\to\infty$, the fields
$\chi_{L,R}$ become auxiliary, implying the following constraints:
\beq
\bar{\vp}_l \ \xi^l_L=0, \qquad \bar{\vp}_l \ \xi^l_R=0\,.
\label{nxiconstraint}
\eeq
%The extra $\xi$ terms in $A_k$ and $\sigma$
%are responsible for the four-fermion part of the Lagrangian in the %gauged formulation.
\begin{center}
 * * *
\end{center}
We finally consider the case of  superpotential different from zero. 
We consider the case of classical coincidence, so that the vortex is BPS and $\phi=\tq=0$. 
Equations for the fermionic zero modes 
$\lambda$ and $\psi_{q \, i}$ are unchanged, and they still provide the fermionic modes $\zeta_R$ and $\xi_R$. Their profile functions are unchanged, and in particular always confined in a radius $\sim 1/\sqrt{\xi}$.

The fermionic zero modes for $\psi$ and $\psi_{\tq \, i}$ are instead modified by the presence of the superpotential;
\bea
 -\frac{i}{g^2} \nablabarslash \psi -
\sqrt{2}\bar{{\psi}}_{\tq \, i} {q}_i^{\dagger} &=& \sqrt{2}  \mu  \bar{\psi}
\nonumber \\
 -i\nablabarslash
{\psi}_{\tq\, i}-\sqrt{2}q_i^{\dagger}\bar{\psi}&=& 0 \ .
\label{nonsusyfermions}
\eea
However, they still provide fermionic zero modes.
Due to the index theorem, we can immediately infer that these zero modes, although modified, must still be present and considered in the effective action. 
Solutions for these zero modes have been studied in \cite{Shifman:2008wv} in the two limits of $\mu$ deformation very small or very large, with respect to $\sqrt{\xi}$. 
We can use an expansion in powers of $\mu/\sqrt{\xi}$ and have
$ \psi = (\psi_L^{\0},\psi_R^{\1})$ and
$ \psi_{\tq} = (\psi_{\tq \,L}^{\1} , \psi_{\tq \,R}^{\0} )$.

%\texttt{ANO vortex with superpotential, why 0,2 is inevitable}

The worldsheet ${\cal N}=(0,2)$ supersymmetry is generated by the two right-moving supercharges $Q^{1}_R$, while the left-moving $Q^{1}_L$ supersymmetries are explicitly broken by the vortex solution. The extended supercharge $Q^2 =(Q^{2}_L,Q^{2}_R)$ are instead broken by the superpotential $W(\Phi)$. 
In the absence of the superpotential, the $\N=(2,2)$ supersymmetry on the wordsheet is generated by $(Q^{2}_L,Q^{1}_R)$. The fact that the left and right generators are orthogonal in the $\SU(2)_R$ space  is clearly visible from the fact that the left and right fermionic zero models come  from totally disconnected Dirac equations, the ones  for $\psi,{\psi}_{\tq\, i}$ and the ones for $\lambda,{\psi}_{q\, i}$.
That is also the basic reason why, when $\N=2$ is broken to $\N=1$, the supercharges on the worldsheet are not $\N=(1,1)$ but $\N=(0,2)$ or $\N=(2,0)$.
An important feature of $(0,2)$ theories is the existence of a fermionic
multiplet $\Gamma$, containing only left-moving fermions $\chi_L$
and no propagating bosons. The fermions can
live in any representation of the gauge group, like chiral multiplets, and, in particular, it is possible introduce a superpotential $J(\Phi)$, a function of the chiral superfields, for each
Fermi multiplet $\Gamma$ \cite{Witten:1993yc}.

Edalati and Tong (ET) proposed the \ntwoo worldsheet deformation induced by the superpotential $W(\Phi)$. In their construction, the \ntwot model (\ref{cpg}), (\ref{cpgf})
is supplemented by the following deformation
\beq
\delta S_{\rm CP \  het} =   \int d^2 x\,
\left\{ - \left|  {W}^{\p}_{1+1}(\sigma) \right|^2 + \left[ \, \zeta_{L}\,
{W}^{''}_{1+1}(\sigma)\,\bar{\chi}_{R}   + {\rm h.c.} \right] \right\} \ ,
\label{etong}
\eeq
breaking \ntwot down to \ntwoo. 
Here ${W}_{1+1}\p$ enters as a  two-dimensional superpotential for the fermionic superfields containing the $\zeta_{R}$ fermions, and is a function of the chiral superfield containing $\sigma$.
Integrating out the axillary field $\chi$ now leads us to the fermion constraints
\beq
\bar{\vp}_l \,\xi^l_L= {\bar{W}}^{''}_{1+1}(\sigma) \,\bar{\zeta}_L, \qquad \bar{\vp}_l \,\xi^l_R = 0 \  .\label{modnxiconstraint}
\eeq
The right-handed fermion remains intact, while the left-handed fermion no longer decouples from 
the orientational one. 
The right-handed fermion $\zeta_R$ as well as the translation modulus $z$
remain free fields.

The general structure of the deformation
(\ref{etong}) follows the \ntwoo supersymmetry.
ET suggested  that the bulk and worldsheet
deformation superpotentials should essentially coincide
\beq
{W}_{1+1} \propto \frac{{W}_{3+1}}{\sqrt{ \xi}} \ ,
\label{etconjecture}
\eeq
up to some constant normalization factor.
The parameter $\sqrt{\xi}$ is essential in order to map a four-dimensional superpotential, which has dimension three, to a two-dimensional one, which has dimension two. And $\xi$ is exactly what physically enables us to map a four-dimensional theory to a two-dimensional one through the Higgs effect, and the creation of vortices.
The analysis of \cite{Shifman:2008wv}
confirms Eq.~(\ref{etconjecture}) at small $\mu$.\footnote{If $\mu$ becomes greater than $\sqrt{\xi}$, other corrections coming from the normalization of the kitetic terms must be taken into account.}
The exact coefficient of proportionality can be evaluated by the fermionic zero mode overlap between $\xi_L$ and $\zeta_L$ (\ref{modnxiconstraint}). 
This computation has a tricky    aspect. 
The solution for the zero modes can be found in the leading order in $\mu/\sqrt{\xi}$. 
Still, we can separate the solutions that are orthogonal ($\sim \bar\vp^a \widetilde{\xi}_{R\,i}$) from the ones that are parallel ($\sim \bar\vp^a \vp_i \zeta_R$) to the vortex orientation. 
The mass $\mu$ does not change this feature.
The  point is that the fermion $\widetilde{\xi}$, when confronted with the ET action, corresponds to the shifted
\beq
\widetilde{\xi}_{L\,i}  =     \xi_{L\,i}  - \frac{\vp_i}{\beta} {\bar{W}}^{''}_{1+1}(\sigma) \,\bar{\zeta}_L \ .
\eeq 
The kinetic term, which is kanonical in terms of $\zeta_L$ and $\xi_{L \, i}$, gives now the cubic interaction \cite{Shifman:2008wv}
\beq
-i \frac{|{W}^{''}_{1+1}|^2}{\beta}
\nabla_R \bar{\vp}_i \, \zeta_L \, \widetilde{\xi}_{L \, i }  + {\rm h.c.}
\eeq 
We then have  to substitute the explicit  zero modes solution and compute the coefficient in front of the interaction. This interaction comes out of the fermions kinetic terms, in the bulk action.
This is the strategy that was pursued in \cite{Bolokhov:2009sg}.

In the large $\mu$ limit, namely  $\mu \gg \sqrt{\xi}$, there is a transition to a different regime.  Since the mass of the $\psi$ fermion becomes very large, we can neglect its kinetic term in (\ref{nonsusyfermions}):
 and rewrite it in a simpler form
\bea
& \bar\psi = - \frac{  {q}_i^{\dagger}}{\mu} {\bar{\psi}}_{\tq \, i} 
\quad {\rm and }\quad
{{\psi}}_{\tq \, R\, i} =  i \bar{{\psi}}_{\tq \,L\, i} &\nonumber \\ [1mm]
&
  -\nabla_z
{\psi}_{\tq\,R\, i} 
+ \sqrt{2}  \frac{q_i^{\dagger} {q}_l^{\dagger}}{\mu} {\psi}_{\tq\,R\, i} = 0 &\ .
\eea
We can see from the second equation that the ${\psi}_{\tq\, i}$ fermion behaves at large distance as $e^{- m_{\rm l} r}/\sqrt{r}$, in accordance with the light mass of (\ref{masseshl}).
That means that, while the fermionic modes $\zeta_R$ and $\xi_{l\ L}$ remain confined in a region $\sim 1/\sqrt{\xi}$, the left-handed ones become spread over a radius $\sim \mu/\xi$. In the limit $\mu \to \infty$, their behavior is the power law $\sim 1/\sqrt{r}$ and thus they become non-normalizable.

%% file: a.tex
Here we consider the corner (A), that is the perturbed $\N=2$ region. 
Here the superpotential and the FI parameter  are very small compared to the dynamical scale $\Lambda_{\snd}$. 
We can thus use the known results about the $\N=2$ strong dynamics, mostly obtained with the Seiberg-Witten technique, and then the perturbations in the low-energy effective Lagrangian.

The SW curve for $n_c=n_f=n$ is the hyperelliptic surface defined by
\bea
{y}^2 &=& {\cal P}_{(n,n)}(z) \nonumber \\
&=& \frac{1}{4} \det(z-\phi)^2-  \Lambda^{n}z^{n} \ ,
\eea\
where the values of $\phi$ label the point in the moduli space of vacua. Massive deformations, and supersymmetry breaking terms, lift the moduli space, leaving, in general, only a discrete number of vacua.   In the case of interest, due to the presence of the FI term, only one vacuum survives: the so-called root of the baryonic branch.\footnote{Technically, there is no baryonic branch, since we are working with $\U(n)$ and not $\SU(n)$. But we still use the name ``root of the baryonic branch'' for simplicity. Otherwise, more correctly, we should call it: ``root of the would-be baryonic branch, if it would not be for the $\U(1)$ F-term.'' But this name is too long.}

The root of the baryonic branch is located at the $\Z_n$ invariant vacuum
\beq
\phi= e^{i\pi/n} {\rm diag} (\Lambda, \Lambda \omega_{n}, \Lambda \omega_{n}^2 , \dots, \Lambda \omega_{n}^{n-1}) \ ,
\eeq
where $\omega_{n}$ is the $n$th root of unity $e^{i2\pi/n}$.
The factorization of the curve in this point is given by the following algebraic steps
\bea
{\cal P}_{(n,n)}(z)  &=& \frac{1}{4} \prod_{j=1}^{n} (z + e^{i\pi/n }\Lambda \omega_{n}^j)^2 - \Lambda^{n}z^{n}    \nonumber \\
&=& \frac{1}{4} \left(z^{n} + \Lambda^{n} \right)^2-  \Lambda^{n}z^{n} \nonumber\\
&=& \frac{1}{4} \left(z^{n} - \Lambda^{n} \right)^2 \ .
\label{factorizationbb}
\eea
Note the peculiarity that all the roots are doubled here.  Away from $u_1 =0$, the singularity splits into $n$ different branches. 
The root of the baryonic branch  is essentially the quantum generalization of the concept of color-flavor locking.
It is not the quantum generalization of the concept of coincidence vacuum \cite{Bolognesi:2008sw}.\footnote{Classically, instead, the origin of the moduli space, $\phi =0$, carries both properties, coincidence and color-flavor locking. 
Coincidence vacua are always lifted by the presence of an FI term, and are, in general, meta-stable vacua.}

Particles and charges can be put in a diagonal form, and are now given in the following table  
\begin{center}
\begin{tabular}{c|ccccccccc}
 && $\U(1)_1$ & $\times$ &  $\U(1)_2$ & $\times$ & $\cdots$ &   $\U(1)_{n -1}$
 & $\times$ &  $\U(1)_{n}$ \\ \hline 
 $E_{1 \phantom{-1}} $    &&   $1$  & &   &      && &&      \\
 $E_{2 \phantom{-1}} $    &&     & & $ 1$  &      && &&      \\
 $\vdots \phantom{....}$  &&& &   & &$\ddots$ &&&   \\
 $E_{n -1}$&&             & &&&  &  $1 $     &     &  \\
 $E_{n \phantom{-1}}$&  &  &  &     & &    &  &       & $1$     \\ \hline
 \end{tabular}
\end{center}
where $E_j$ are the various charged hypermultiplets, accompanied by the  partners $\tE_j$ with the opposite charge. 
%\texttt{introduci i nomi e le definizioni dei campi A, E, ...}

Now let us examine SUSY breaking (from $\N=2$ down to $\N=1$) in the effective low-energy theory on this vacuum, after introduction of a superpotential like (\ref{supsenzalin}).
In this case, the low-energy superpotential is
\beq
{\cal W}   = \sqrt{2} \left(  \sum_{j=1}^{n} \tE_j A_j E_j  + \mu u_2(A_1, \dots, A_{n} )   \right) \ .
\eeq
The potential is then
\beq
V = \sum_{j=1}^n \left( 2g_j^2\left|\te_j e_j + \mu \frac{\partial u_2}{\partial a_j}\right|^2+\frac{g_j^2}{2}(|e^j|^2-|\te_j|^2-2\xi)^2 \right) \ .
\eeq
Each $\U(1)_j$ is Higgsed by the condensation of the respective hypermultiplet $E_j$, $\tE_j$, and each admits a formation of an ANO vortex. 
To compute the tension of these vortices we can use the technique described in $(\ref{newv})$, where $-\Re W'$, $\Im W'$, and $\xi$ are considered as a triplet of the $\SU(2)_R$ symmetry.

To compute the $\te_j e_j$ condensate, from the vanishing of the $F_{A_j}$ term, it is more convenient to invert the matrix relationship and write it as
\beq
\label{matrice}
\sum_{j=1}^{n}  \te_j e_j \frac{\partial a_j}{\partial u_l}  = -  w_l \ ,
\eeq
where $w= (0,\mu,0,\dots,0)$ is the vector of coefficients of the superpotential.
We can then use the SW solution for the $\partial a_j / \partial u_l$ as period integrals of the holomorphic differentials,
\bea
\frac{\partial a_j}{\partial u_l} &=& \frac{1}{2\pi i } \oint_{\alpha_j} \frac{z^{n-l} dz}{z^n -\Lambda^n} \nonumber \\
 &=& \frac{1}{\Lambda^{l-1}} \frac{{{\omega}_n}^{j(n-l)}}{ \prod_{k\neq j}^n ({{\omega_n}}^j -{{\omega_n}}^k)} = \frac{1}{\Lambda^{l-1}} 
\frac{{\omega_n}^{j(1-l)}}{ \prod_{k=1}^{n-1} (1 -{\omega_n}^k)} \nonumber \\ [2 mm]
&=&
\frac{{\omega_n}^{j(1-l)}}{n \Lambda^{l-1}} \ .
\eea 
The solution of Eq.~(\ref{matrice}) is then given by
\beq
\te_j e_j = - \mu \Lambda \  {\omega_n}^{j} \ .
\eeq
For every $l \neq 2$, the sum $\ref{matrice}$ vanishes due to the complex phases. Only for $l=2$ do the phases cancel precisely and we get $-\mu$.

The $n$ vortices have degenerate tension:
\bea
T_j &=& 4\pi \sqrt{\left|\mu \Lambda e^{ \frac{i2\pi j}{n}}\right|^2 +\xi^2} \nonumber \\[1mm]
    &=& 4\pi \sqrt{|\mu \Lambda|^2 +\xi^2} \ .
\label{tensionswregime}
\eea
This computation is valid only in the small $\mu$ limit, where the second derivatives of the superpotential have subleading contribution to the tension. 
But we know that these second derivatives have the effect of making multiple vortices interacting, and, in particular, type I.

As observed in \cite{quantumvortexstring}, and then in \cite{Shifman:2008kj}, the heterotic vortex theory (corner (C) to be discussed in Section \ref{c}), predicts $n$ degenerate ground states for the non-Abelian string, and dynamical breaking of supersymmetry. 
%This phenomenon is pretty much understood fromt he point of view of the $1+1$ theories. When the $\N=(0,2)$ perturbation is added, there is no more {\it Witten %index} to protect the existence of the $n$ SUSY vacua.  Although classically they are still supersymmetric, we do not expect this to be still true when quantum %correction are turned on. The large $n$ limit computation confirms this expectation.
So we find agreement with the finding of the heterotic vortex theory: $n$ degenerate non-BPS vacua/vortices.  The physical interpretation is that the $n$ vortices we observe in the SW low-energy effective action  are nothing but the ground states of the non-Abelian heterotic vortex theory of corner (C).

Now we change the superpotential, and consider the possibility of a linear term of the type (\ref{sup}). The low-energy effective superpotential becomes
\beq
{\cal W}   = \sqrt{2} \left( \sum_{j=1}^{n} \tE_j A_j E_j  + \mu (u_2 -a u_1)  \right) \ .
\eeq
Physics is similar to the previously discussed case. Every $\U(1)_j$ is Higgsed and is accompanied by the formation of an ANO vortex. The only difference is that the $n$ vortices no longer have degenerate tensions.  We can still use   Eq.~(\ref{matrice}),  now with $w= (a,\mu,0,\dots,0)$. The solution is now given by
\beq
\te_j e^j = - \mu \left( \Lambda \  {\omega_n}^{j} -a \right) \ .
\eeq
The term proportional to $a$, when considered in the sum (\ref{matrice}), affects only the component with $l=1$.

The coefficient $a$ presents is in fact an explicit breaking of the $\Z_n$ symmetry.
Something peculiar happens for the specific choice
\beq
a = \Lambda e^{\frac{i 2 \pi k}{n}}  \ .
\eeq
The tension of the vortices is now
\bea
T_j &=& 4\pi \sqrt{\left|\mu \Lambda (e^{\frac{i2\pi j}{n}} -e^{\frac{i 2 \pi k}{n}})\right|^2 +\xi^2} \nonumber \\[2mm]
&=& 4\pi \sqrt{ \left| 2 \mu \Lambda \sin{\pi (j-k)} \right|^2 +\xi^2 } \ ,
\eea
and we can see that the $k$-vortex is BPS saturated, with the tension $T_k =4\pi \xi$.

We could have also used  the following formula to compute the tensions \cite{Bolognesi:2004yh,Auzzi:2004yg}:
\beq
T_j = 4\pi \sqrt{|W\p({\rm root}_j)|^2 + \xi^2} \ ,
\eeq
where ``root$_j$'' is the $j^{\rm th}$ root of the SW curve (\ref{factorizationbb}). This is the quantum generalization of (\ref{newv}). 
Derivation of this formula becomes particularly transparent in the MQCD formulation of the theory. At the root of the baryonic branch, since all the roots are doubled, the MQCD curve is composed by two disconnected parts: one that becomes asymptotically the NS$5$-brane, and another the NS$5'$-brane. Superpotential deformation is obtained classically by giving a certain shape to the NS$5'$-brane. Since in general the two branes are connected, and become in MQCD  a unique embedded Riemann surface, the superpotential acquires quantum corrections by requiring the matching of the two branes. Since at the root of the baryonic branch they are disconnected, the superpotential does not receive quantum corrections. There are though quantum correction to the tension of vortices. They come from the point were $W\p$ should be computed. It must be computed where the distance between the two disconnected branes is minimized, and this is {\it not} at the root of $W\p$, but at the root of the SW curve.

%% file: b.tex
This corner is the one where the strong dynamics is that of $\N=1$ SQCD in $3+1$ dimensions. 
The mass $\mu$ of the adjoint field is very large compared to the dynamical scale $\Lambda_{\snu}$. 
We can thus use the known results about the four-dimensional dynamics, and then consider the FI term as a small perturbation to the last one.
In some respects, what we are going to do  is very similar to what we have done in corner (A). 
The quantum effects are four-dimensional, and treated with the known available techniques. 
The vortices that we observe are discrete solutions, with no internal moduli space. 
We then interpret them as the ground states of the heterotic vortex theory that we obtain in corner (C).

The classical theory has a moduli space of vacua,
parametrized in a gauge invariant way by the meson superfield
\beq M_{ij}=\tQ_i Q_j \ ,\eeq
together with the two baryonic superfields
\beq 
B=\epsilon_{\a_1\ldots \a_{n}}Q^{a_1}_1\ldots
Q^{\a_{n}}_{n} \ , \qquad 
\tB=\epsilon_{\a_1\ldots
\a_{n}}\tilde{Q}^{\a_1}_1\ldots \tilde{Q}^{\a_{n}}_{n} \ .
\eeq
These are not independent and obey the classical constraint
\beq 
\det M_{ij} - B\tB=0 \ .
\label{clascons}
\eeq
The light fields in the classical theory are $M$, $B$, and
$\tilde{B}$ subject to the constraint \eqn{clascons}. 
%We can say that we are in confined phase but without mass gap.
The resulting manifold has singularities due to the massless gauge bosons, which emerge when the symmetry breaking is not maximal.

In   quantum theory, the classical
constraint \eqn{clascons} is modified by the dynamical scale \cite{Intriligator:1995au},
\beq 
\det M_{ij} - B\tB=\Lambda^{2n} \ .
\label{qdef}
\eeq
%where $\Lambda_{\snu}^{2N} = \mu^{2N} e^{-4\pi^2/e^2(\mu)+i\theta}$
%is proportional to the one-instanton action. 
The manifold defined
by \eqn{qdef} is smooth, and the singularities of the classical
moduli space have been resolved.
At an energy scale smaller than $\Lambda$, the degrees of freedom are the massless moduli of the manifold (\ref{qdef}).
The physical interpretation, at the base of the manifold $\tilde{B}=B=0$, is that we have confinement and chiral symmetry breaking.

In our case, since we work in $\U(n)$ and not $\SU(n)$, there is also a residual $\U(1)_B$ gauge interaction, coupled with the baryons, which remains weakly coupled in the infrared. 
In what follows, we want  to consider the deformation with a small FI term $\sqrt{\xi} \ll \Lambda$.
In this approximation, we just need to consider  the mesonic and baryonic massless moduli (at least for the chiral sector).
\begin{center}
 * * *
\end{center}
The situation we are going to face is analogous to the one considered in \cite{Penin:1996si,Achucarro:2001ii,Yung:1999du} regarding vortices in the presence of flat directions.
The vortex core interacts with the extra massless $\tq$ fields. That even if the BPS solution, the ground state, is obtained by setting $\tq=0$.  
Let us consider first the classical case (\ref{clascons}), in order to briefly recall the basic features of vortices in these circumstances. 
The vacuum manifold is parametrized by $q$ and $\tq$ subject to the D-term constraint
\beq
|q|^2-|\tq|^2 = 2\xi \ .
\eeq
Let us call $\tq=0$, $|q|^2=2\xi$ the base of this Higgs branch. 
If we sit at the base, we can have a perfectly well-defined non-Abelian vortex, like (\ref{vortexconf}), while keeping fixed the boundary conditions at infinity.
The situation is completely different when we want to move away from the base of the manifold. 
If we do so, the core of the vortex becomes a source for some of the massless fields that compose  these extra flat directions. 
The outcome is a logarithmic  tail, outside the core of the vortex. This means that is does not make sense to consider a vortex with boundary conditions different from those of the base of the manifold. The logarithmic tail inevitably sets the boundary conditions to infinity.

Now we consider the quantum moduli space (\ref{qdef}).  
As the FI parameter satisfy $\sqrt{\xi} \ll \Lambda$, we can consider it as a deformation to this moduli space. 
The physical interpretation is the following. 
The theory remains four-dimensional all the way down to $\sqrt{\xi}$. 
We can thus use the known results about the strong dynamics of the $\N=1$ SQCD in four dimensions.
The low-energy effective theory contains mesons and baryons, subject to the constraint (\ref{qdef}). 
We then have the FI term that we can introduce at this level of the effective Lagrangian. Its effects are in the D-term of the baryons.
%\beq 
%|B|^2-|\tB|^2 = (2\xi)^n \ .
%\label{dtermbaryon}
%\eeq
The presence of the FI parameter ensures that the
$\U(1)_B$ gauge symmetry is broken,  and this  implies the
existence of strings, similar to the ordinary ANO vortex \cite{quantumvortexstring}.

One could argue that the vortex string we observe in the pure $\N=1$ (large-$\mu$) is just an Abelian vortex, and thus has nothing to do with the non-Abelian vortex we discussed in the small-$\mu$ limit. 
But note that since it is created by the winding of the baryon field $B$, it has $1/n$ the charge of an Abelian $\U(1)_B$ vortex. 
It thus have the {\it same} flux of the non-Abelian string with respect to the $\U(1)_B$. 
Clearly, the string we observe in this limit {\it is} the non-Abelian string, or at least the remnant of it.
Non-Abelian moduli are not visible because they are gaped by the strong dynamics. 
Here, and also in the previous corner (A), they are gaped by the four-dimensional dynamics.

But we also have to discuss the important issue of the flat directions. 
Vortices exist only if
$\tilde{B}=0$, and they are also BPS. 
The quantum mechanical analog of the base of the Higgs branch  is now split into $n$ different $n$ different vacua parametrized by
\beq
\label{meson}
M_{ij}= \mathbf{1}_{ij} \ \Lambda^2 \ e^{\frac{i 2\pi k}{n}} \ , 
\eeq
with $k=1, \dots , n$.
The mesonic field is diagonal, and proportional to the   $n$'th root of the identity.
Whenever we choose to move from one of these $n$ bases, $\tilde{B}\neq 0$, any vortex solution must
necessarily be non-BPS and have the log tail, typical of vortices in the presence of flat directions.
%Our findings, so far, do not match completely the ones of the previous Section. In particular we still do not see the $n$-degeneracy of the vortices, essential in order to interpret them as genuine non-Abelian vortices. 
%In what follows we shall explain why.
\begin{center}
 * * *
\end{center}
To find agreement with the other approaches, we must consider the effect of the operator obtained after integrating out the adjoint field $\Phi$
\beq
\label{theoperator}
W = -  \frac{1}{{\mu \sqrt{2} }} \ \Tr Q \tQ \  Q \tQ \ .
\eeq
This operator, in the low-energy effective theory, is a mass term that lifts the mesonic moduli $M$,  and forces the theory to live in the vacuum where $M=0$.
In the $\mu \to \infty$ limit, the operator clearly vanishes, but the information we are looking for (degeneracy and susy-breaking of vortices) are all encoded in this $1/\mu$ effect, and thus it is important to keep track of this term.

Vortices are non-BPS, and this is due to the quantum deformation of the moduli space. The effective superpotential (\ref{theoperator}) forces the vacuum to be at $\det M=0$, and this implies $B \tB=-\Lambda^{2n}$. The size of this vortex is $\propto \mu/\Lambda^2$, and in particular they become  infinity spread in the $\mu \to \infty$ limit. 
The vortex must have the following profile structure, in the adiabatic limit $\mu \gg \Lambda$:
\beq
B= e^{i \theta} b(r) \ , \qquad \tB = e^{-i\theta} \tb(r) \ ,
\eeq
and for the meson
\beq
M_{ij} =  e^{\frac{i k 2 \pi}{n}} (b\tb +\Lambda^{2n})^{\frac{1}{n}} {\bf 1}_{ij} \ .
\label{degeneracy}
\eeq
Note an important distinction with the semiclassical case (\ref{vortexconf}). Classically, the vortex is obtained by the winding of one of the $n$ quarks, while the others are just spectators. Clearly, the choice of the quark that makes the winding is arbitrary, and thus the emergence of the classical moduli space. Every configuration, though, corresponds to a particular choice of the quark, and thus the symmetry is broken down to $\U(1) \times \SU(n-1)$. In the case under consideration, the symmetry is not broken (\ref{degeneracy}), and we have only a discrete $n$-fold degeneracy. That is exactly what is obtained by an analysis of corner (C) with the worldsheet effective action.

The degeneracy is {\it not} visible from the boundary conditions at $r \to \infty$. It is encoded in the value of the $M$ field at the core of the vortex.
We thus find agreement with the other regions of parameters (A) and (C). 
We have $n$ degenerate non-BPS vortices. 
The degeneracy is {\it not} visible from the boundary conditions at $r \to \infty$. 
The fact that in the $\mu \to \infty$ limit the vortices are infinity spread  is not in contradiction with our previous result. 
We in fact {\it do not} expect quantitative agreement between the various approaches.

%%%%%%%%%%%%%%  critical superpotential
In the case of generic linear term in the superpotential,
%\beq
%\mu \Tr \left( \frac{\Phi^2}{2} - a\Lambda_2 \Phi \right)
%\eeq
the effective generated superpotential is
\bea
W  &=& -  \frac{1}{{\mu \sqrt{2} }} \Tr (Q \tQ \  Q \tQ)  + \sqrt{2} a    \Tr Q \tQ  \nonumber \\[3mm]
&=& - \frac{1}{{\mu \sqrt{2} }} \left[ \Tr (Q \tQ \  Q \tQ)  - 2 a \mu \Tr Q \tQ  \right] \ .
\eea
where we have used the relationship (\ref{relationofscales}) between the dynamical scales. Note that $a$ does not scale with $\mu$.
If we want to have vanishing $\tB$, and consequently a BPS vortex, the  meson  must be in one of the $n$ bases (\ref{meson}). 
This is obtained by choosing
\beq
a=  \frac{\Lambda_{\snu}^2}{\mu} e^{\frac{i2 \pi k}{n}} = \Lambda_{\snd} e^{\frac{i2 \pi k}{n}}\ .
\eeq
For this particular superpotential, we have a BPS string.
\begin{center}
 * * *
\end{center} 
If we want to perform some concrete computation with these kinds of vortices, we need some toy model that captures the essential properties we previously discussed.
Vortices are created by the condensation of the baryon, but due to its dimension, it is not clear how to use it in an effective low-energy theory coupled to the gauge boson $\U(1)_B$. The solution to this problem, would require   knowledge beyond the chiral sector of the theory.

What we will study for now is a simplified model, which essentially corresponds to the extreme case $n=1$. We can thus say that $B=Q$, $\tB=\tQ$ and the quantum deformation (\ref{qdef}) is $M=\tQ Q +\Lambda^{2}$ 
We also used the adiabatic approximation, that is, we impose the condition of the quantum-deformed manifold (\ref{qdef}) over all the vortex profile. 
That is a good approximation in the $\Lambda \gg \mu$ limit since, as we shall see, the field variations become slow enough to consider the constraint (\ref{qdef}) valid at any radius.
We can thus rewrite the superpotential (\ref{theoperator}) for the mesonic field, in terms of the quark
\beq
\label{superpottoy}
W =  - \frac{1}{{\mu \sqrt{2} }} \ \left( \tQ Q + {\Lambda}^2  \right)^{2}   \ .
\eeq
The potential for the constituent quark $q$ is thus
\beq
V= \frac{1}{ 2 \mu^2} \left(\tq q  + {\Lambda}^2 \right)^2 ( |q|^2 +|\tq|^2 )+ \frac{g^2}{2}   \left(|q|^2-|\tq|^2-2\xi\right)^2  \ .
\label{potenziale}
\eeq

The meson mass term stabilizes the log tail typical of vortices with flat directions. 
Let us call $r$ the radial direction out of the vortex center. 
First there is a core consisting of the BPS vortex. 
The core has width $1/\sqrt{\xi}$. 
The $q$ field goes from zero to $\sqrt{\xi}$, $\tq$ remains zero, and the meson is equal to one of the $n$ values of (\ref{meson}). 
There is then a tail where both $q$ and $\tq$ become  of order $\Lambda$ and the meson field goes to zero. 
As $\mu \to \infty$, these vortices become  infinitely spread.
The ansatz for the profile functions is as usual:
\beq
A_k=-\epsilon_{kl}\frac{\hat{r}_l}{r}f(r) \ , \qquad q = e^{i\theta}\sqrt{2{\xi}} \, q(r) \ , \qquad \tq = e^{-i \theta}\sqrt{2{\xi}} \, \tq(r) \ .
\eeq

\begin{figure}[h!]
\epsfxsize=8cm
\centerline{\epsfbox{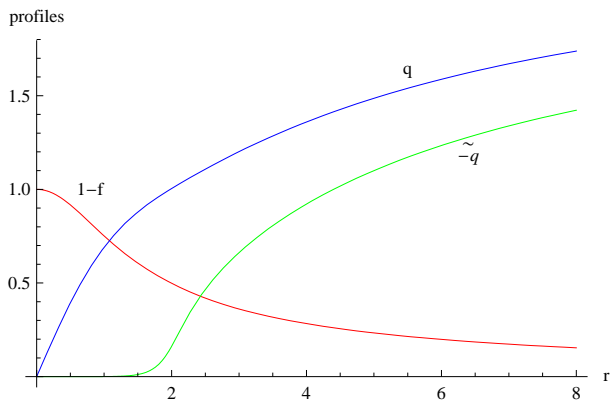}}
\caption{{\footnotesize The vortex core, with the scalar fields profiles and the $\U(1)_B$ gauge field.}}
\label{vortice}
\end{figure}

\begin{figure}[h!t]
\epsfxsize=8cm
\centerline{\epsfbox{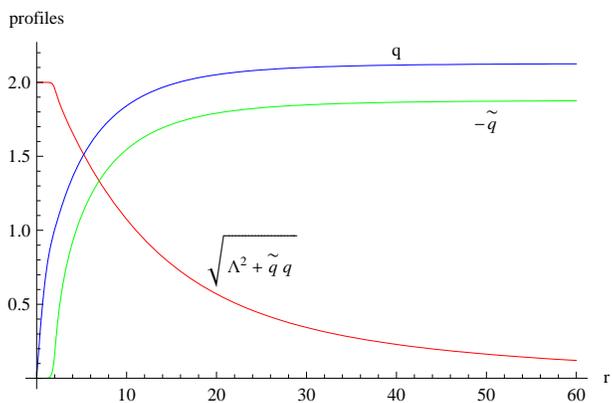}}
\caption{{\footnotesize The profile functions for $q$, $\tq$ at bigger scale.}}
\label{profili}
\end{figure}

\begin{figure}[h!]
\epsfxsize=8cm
\centerline{\epsfbox{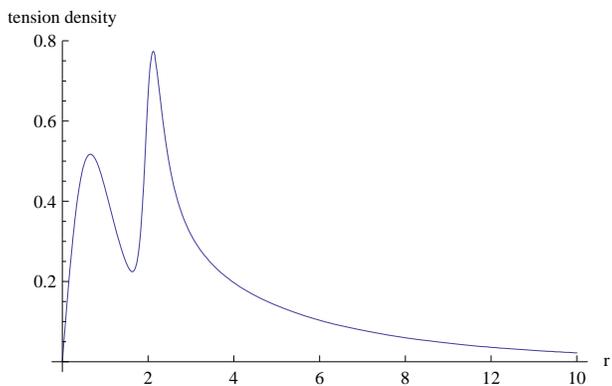}}
\caption{{\footnotesize The tension density, with the two peaks. }}
\label{tensione}
\end{figure}

We present some numerical results for the vortex obtained with the potential (\ref{potenziale}), with parameters $e=1$, $\xi =.5$, $\Lambda=2$ and $\mu =100$. 
The first two plots are some profile functions for $q$ and $\tq$. Figure \ref{vortice} shows the core of the vortex, with also the gauge field profile $1-f$ (as defined in (\ref{vortexconf})). 
Figure \ref{profili} is on a larger scale, where it is visible that the tails of  $q$ and $\tq$ saturate  to their vacuum expectation. Here we also plot $\sqrt{\Lambda^2 +\tq q}$, which is related to the mesonic condensate. Figure \ref{tensione} shows the tension density.
The first peak corresponds to the ANO vortex in the core. The second peak starts when the $\tq$ field starts to grow.

We can make a comparison with the tension formula obtained in the SW regime (\ref{tensionswregime}). If we take this formula and extrapolate to the limit under consideration, we get
\beq
T = 4 \pi \Lambda_{\snu}^2  +  4 \pi \xi + \dots
\eeq
The second term $4\pi \xi^2$ corresponds to the tension of the BPS vortex in the core, roughly the first peak in the tension profile of Figure \ref{tensione}.  The first term, the biggest one, should correspond to the second peak. But here, the formula (\ref{tensionswregime}) gives a result that is, in general,  far from the real value.

We need now to distinguish two regimes inside the corner (B). We need to compare the mass of the light scalar field   $\sqrt{2} \Lambda^2 / \mu$ with the mass of the photon  $m_{\gamma} = e \sqrt{2} \Lambda$.\footnote{Since $\U(1)_B$ is infrared free, the coupling is $e^{-2} \sim \log(\Lambda/\Lambda_{\rm cutoff})$ with $\Lambda_{\rm cutoff}$ the high-energy scale where $\SU(n)$ and $\U(1)$ are unified.}. 
If 
\beq
\frac{\Lambda}{\mu} \ll  e 
\label{primacond}
\eeq
we can for the moment ignore  $\xi$ in the potential (\ref{potenziale}), we can simplify it by setting
$q=-\tq$, and also rescaling by a $\sqrt{2}$ factor in order to have just one field $q$ with the simplified potential
\beq
V= \frac{1}{ 8 \mu^2} |q|^2 \left( |q|^2  - 2 {\Lambda}^2 \right)^2  \ .
\label{potenzialesemplice}
\eeq
The prediction $4\pi \Lambda^2$ is equivalent to the tension of a BPS vortex with the same condensate given in (\ref{potenzialesemplice}).
The SW formula is expected to capture in some sense the BPS aspect, and in particular it fails in this region of parameter. If the conditions of corner (D) plus the condition (\ref{primacond}) satisfied, these vortices are strongly type I. The tension is smaller than $4 \pi \Lambda^2$ \cite{Yung:1999du}, and gets logarithmically suppressed $ \sim .4 \pi \Lambda^2 / \log(\mu e /\Lambda)$. 
That means that the tension of the vortex is essentially dominated by
the BPS core $T \sim 4 \pi \xi$ plus small correction. This result is mirrored by what we shall find in the corner (D) of the quantum heterotic vortex.  Also here, for a reason also related to the flat directions, the quantum corrections to the BPS tension, are vanishing (see (\ref{suppression})).

In the other case $\Lambda/\mu \gg e$ the tension is now enhanced due to the fact the up to the radius $1/m_{\gamma}$ they are essentially global vortices.  The tension, in this case, is logarithmically enhanced: $T \sim 4 \pi \Lambda^2 \log(\Lambda/(\mu e))$.

%\footnote{Note that since thm_{\gamma}e $\U(1)_B$ is infrared free, we have $1/e^2 \sim \log (\Lambda/\sqrt{\xi})$. So there are particular regions, inside corner (D), in which $m_{\gamma} \ll m_q$ and the vortex becomes type II. We have not considered those in detail. For this, we do not think that being the vortices type I is a generic statement over all the parameter space.}

%Compare $2 e^2$ wth $  |q|^2 / \mu^2$
%\beq
% \log  \frac{\sqrt{\xi}}{\Lambda} \ \  ? \ \ 
%\frac{\mu^2}{\Lambda^2} 
%\eeq

%It is oppotune to discuss the relation with the work \cite{} and the coincidence vacua. So we do not have discussed them, and the reader may wonder why they do %not play any role in this discussion. The reason is that the FI term. lifts all the vacua with the exception of the root of the barionic branch.  

%% file: c.tex
Now we deal with region (C), that of the heterotic vortex theory. We call it in this way because as $\sqrt{\xi} \gg \Lambda$, and $\mu$ still sufficiently small, all the $3+1$ degrees of freedom are gaped in the bulk, and the  nontrivial dynamics happen  only on the $1+1$ worldsheet. Due to the internal orientational  moduli, the dynamic is highly non-trivial. The coupling runs and becomes strong in the infrared.
We can study the vacuum structure and mass spectrum in the large $n$ limit, following \cite{Shifman:2008kj}. 
The technique is the basic one used in \cite{Witten:1978bc,D'Adda:1978uc} to solve the non-supersymmetric and the $\N=(2,2)$ $\CP(n-1)$ model.

The new thing, with respect to \cite{Shifman:2008kj}, is that we want to consider the presence of a generic linear term in the superpotential
\beq
\label{generic}
W_{1+1} = \omega \left( \frac{\sigma^2}{2} -a \sigma \right) 
\eeq
We used the dimensionless coupling $\omega$ defined by
\beq
\omega \propto \frac{\mu}{\sqrt{\xi}} \ .
\eeq
For the superpotential $\mu \Tr \phi^2/2$, we can just quote the results of \cite{Shifman:2008kj}, and see that they confirm the previous analysis of corners (A) and (B).
We want now to perform the large-$n$ limit in the case of a generic superpotential (\ref{generic}).

The heterotic \ntwoo model, in the gauged formulation, consists of the bosonic part (\ref{cpg}), plus the fermionic (\ref{cpgf}), plus the heterotic deformation (\ref{etong}) which becomes
\beq
\delta S_{\rm CP het}
= \int d^2 x 
\left\{-
|\omega|^2\,\left|\sigma - a
\right|^2 +
\left[\   \omega \,\zeta_L \bar{\chi}_{R}\, + {\rm h.c.}\right]
\right\} \ .
\label{cpg02}
\eeq
%where we omitted the fields  $z$ and $\zeta_L$
%since they are free and totally decoupled from the previous ones. 
%This  action is the one obtained in \cite{Edalati:2007vk}.
Note that the $a$ parameter enters only in the potential for the $\sigma$ field. 
The $\mu$ parameter enters in the potential for $\sigma$, and also in the coupling between the $\chi_L$ and $\zeta_R$. 
Integrating over the axillary fields $\chi$, we arrive at the constraints (\ref{modnxiconstraint}):
\beq
\bar{\vp}^l \,\xi^l_L= \omega^* \,\bar{\zeta}_L, \qquad \bar{\vp}^l \,\xi^l_R=0 \ .
\label{modconstraint}
\eeq

We need first to establish the  $n$-scaling of the various parameters in the action:
\beq
\beta_0 \sim n \ , \quad \Lambda_{3+1} \sim n^0 \ , \quad \mu,\ \xi \sim n \ .
\eeq 
The gauge coupling $g^2$ scales like $1/n$ and that imply that the radius of the $\CP(n-1)$ manifold scales like $\beta_0 \sim n$. The dynamical scale $\Lambda_{\rm CP}$ is thus constant. 
The scaling of $\mu$ and $\xi$ is in order to have all the physical scales $m_{\rm h}$ and  $m_{\rm l}$ in (\ref{masseshl})  to scale like $n^0$.
From now on, in this section, we use $\Lambda$ to refer to the dynamical scale of the CP model.

In the gauge formulation, the Lagrangian is quadratic in the fields $\vp$ and $\xi$, and all the interactions happens through the intervention of the auxiliary fields.  We can thus integrate them out, just computing the determinant of the quadratic form. 
%We are thus left with the %following quanutm effective action
\beq
\left[
\frac{{\rm det}\, \left(-\pt_k^2 
-2|\sigma|^2\right)}{ {\rm det}\, \left(-\pt_{k}^2 +D
-2|\sigma|^2\right)}
\right]^n \ .
\eeq
The first determinant comes from the
boson loops while the second comes  from the fermion loops. 
Note that the $\vp^{l}$ mass 
 is given by $2|\sigma|^2-D$,  while the fermion $\xi^l$
mass is $2|\sigma|^2$. 
The $D$ field must vanish in order to have the supersymmetry unbroken.

Let us start as in \cite{Witten:1978bc} by first evaluating the expectation value for the auxiliary field $D$.
\beq
i \beta + n \int \frac{d^2k}{4\pi^2} \frac{1}{k^2 + D  -2|\sigma|^2 + i\epsilon} = 0
\eeq
We regularize the divergence by a rigid cutoff $\murg$ in the momentum space, and the result is
\beq
 2|\sigma|^2 - D  =  \murg^2\  e^{-4\pi \beta / n} = \Lambda^2
\eeq
with $\Lambda$ the dynamical scale defined to be invariant under the change of the cutoff scale. That also defines the renormalization of the coupling as:
% defined by
%\beq
%\Lambda = \murg e^{-2 \pi \beta}
%\eeq
\beq
\beta(\murg) = \frac{2 \pi }{g(\murg)^2} = \frac{n}{4\pi} \log \left( \frac{\murg}{\Lambda} \right)
\eeq

%The effective potential as 
%a function of the $D$ and $\sigma$ fields
%\beqn
% V_{\rm eff}(D,\sigma) &=& \int d^2 x \,\,\frac{n}{8\pi}\,
%\left\{
% -\left(D+2|\sigma|^2\right)
%\log\, {\frac{D+2|\sigma|^2}{\Lambda^2}} +D\right.
%\nonumber\\[3mm]
%&&  +   \left.
%2|\sigma|^2\,\log\, {\frac{2|\sigma|^2}{\Lambda^2}}
%+ \frac{ 8 \pi |\omega|^2}{ n} |\sigma-a|^2   \right\}\,,
%\label{effpot}
%\eeqn
%Minimizing this potential with respect to $D$ we get
%\beq
%\beta_{\rm ren}=\frac{n}{4\pi}\, 
%\log\, {\frac{D +2|\sigma|^2}{\Lambda^2}}=0
%\label{veq}
%\eeq
%whose solution is 
%\beq
% D +2|\sigma|^2 =\Lambda^2 \ .
%\label{sol}
%\eeq
%Since $D$ is auxiliary (it cannot get a kinetic term, not even through quantum corrections),
% we want to integrate it out 
The effective potential for the field $\sigma$  is
\beq
 V_{\rm eff}(\sigma)=\frac{n}{8\pi}\left\{
\Lambda^2 + 2|\sigma|^2  \left(\log\, {\frac{2|\sigma|^2}{\Lambda^2}} -1 \right)
+ 8 |\sigma-a|^2\,u\,     \right\}.
\label{sigmapot}
\eeq
where instead of the deformation parameter $\omega$ we introduced the dimensionless parameter $u$, which does not scale with
$n$:
\beq
u=\frac{   \pi |\omega|^2}{   n}  \ .
\label{u}
\eeq
The minimum is given to the solution of the following equation:
\beq
\frac{2 |\sigma|^2}{\Lambda^2} = \exp{\left(-4 u \frac{\sigma-a}{\sigma}\right)} \ .
\label{minimization}
\eeq
To have a supersymmetric solution, we need $D=0$ and  to solve (\ref{minimization}). There are only two possibilities for that to happen:
\begin{itemize}
\item $ \ \mu =0 \ $ This is the trivial $\N=(2,2)$ case without the superpotential;
\item $ a=\frac{\Lambda}{\sqrt{2}} \ $  This is the non-trivial case we were looking for. A particular tuning of the linear term  in the superpotential preserves supersymmetry. 
\end{itemize}
The next part will be devoted to the last particular case, and to solve for the spectrum of the theory.

Before that, we want to consider the $a=0$ case, when supersymmetry is broken, and make a quantitative comparison with the tension formula obtained in the SW regime (\ref{tensionswregime}). If we take this formula and extrapolate to the limit under consideration, we get
\beq
T = 4 \pi \xi  + 2\pi \frac{|\mu \Lambda_{3+1}|^2}{\xi} + \dots
\eeq
Clearly this is an extrapolation, behind the limit of validity of corner (A). But we want nevertheless to make a comparison. The quantum effective potential for $\sigma$ is (\ref{sigmapot}) with $a=0$. the minimum is at 
$\sigma=\Lambda e^{-2u}/\sqrt{2}$, and the vacuum energy density is thus
\beq
V=\frac{n \Lambda^2}{4\pi} \left( 1-e^{-4u} \right) = \Lambda^2\omega^2 + \dots \ .
\eeq
This time the formula of corner (A) correctly reproduces the tension; at least up to a proportionality factor, the parametric dependence upon $\Lambda$, $\mu$ and $\xi$ is correct.\footnote{We have not yet explicitly checked if also the numerical factor is correct. To do so, we should confront   the proportionality factor in (\ref{etconjecture}) that, up to some numerical integral, has been computed in \cite{Shifman:2008wv}.}
\begin{center}
 * * *
\end{center}
We consider the effective Lagrangian for the fields $A_k$, $\sigma$, and $\chi_{L,R}$. These fields acquire a kinetic term through quantum corrections.
In the large-$n$ limit, the kinetic terms and the couplings can just be computed evaluating the $1$-loop diagrams with the $n$ and $\xi$ fields running inside the loops. 
The effective action is
\bea
S_{\rm eff}
&=&
 \int d^2 x \left\{
-\frac1{4e_{\gamma}^2}F^2_{kl} + \frac2{e_{\sigma}^2}
|\pt_k\sigma|^2
\right.
\nonumber\\[3mm]
 && + 
\frac1{e^2_{\chi}}\,\bar{\chi}_{R}\,i\,\pt_{L}\,  \chi_R
+\frac1{e^2_{\chi}}\,\bar{\chi}_{L}\,i\, \pt_{R}\, \chi_L
+  \bar{\zeta}_L \, i
\, \pt_R \, \zeta_L
\nonumber\\[3mm]
 && 
\left.
-V(\sigma)+  \frac{n}{2  \pi}\,\frac{{\rm Im}\,\sigma}{|\sigma|}\,\widetilde{F}
  - \left[ \sqrt{2}\,\Gamma\,\sigma\,\bar{\chi}_{R}\chi_L  
- \omega \,\bar{\chi}_{R}\,
\zeta_L  +{\rm h.c.}\right] \right\},
\label{effectivection}
\eea
where 
$V(\sigma)$ is given in Eq.~(\ref{sigmapot}), and $\widetilde{F}$ is the  dual gauge field strength $\widetilde{F}=\frac12\varepsilon_{kj}F_{kj}$.
Here $e^2_{\gamma}$, $e^2_{\sigma}$, and $e^2_{\chi}$ are the coupling constants that
determine the wave function renormalization for  the photon, $\sigma$,
and $\chi$ fields. 
$\Gamma$ is the induced Yukawa coupling. 
These couplings are given by one-loop graphs which we have been computed in \cite{Shifman:2008kj}.
For the case in which we are interested, $D=0$ and $a=\Lambda/\sqrt{2}$, they are
\beq
\frac1{e^2_{\sigma}}=\frac1{e^2_{\gamma}}=\frac1{e^2_{\chi}}=
\frac{n}{4\pi}\,\frac{1}{2|\sigma|^2}=\frac{n}{4\pi}\,\frac{1}{\Lambda^2}\,,
\label{e22}
\eeq
while the Yukawa coupling  is
\beq
\Gamma=\frac{n}{4\pi}\,\frac{2}{\Lambda^2}\ ,
\label{Gamma22}
\eeq
Since $D=0$, they are exactly the same as the ones of the $\N=(2,2)$ theory.

The $({\rm Im}\,\sigma)\, \widetilde{F}$ mixing was calculated in \cite{Witten:1978bc}
for \ntwot theory. This mixing  is due to
the chiral anomaly and is the term that gives mass to the photon in the effective Lagrangian. 
The $\U(1)$ rotation of the field $\sigma$ is broken by the anomaly down to $\Z_n$.
Since the anomaly is not modified by the superpotential  deformation, we can use 
the same result in the deformed $\N=(0,2)$ theory.

In order to compute the spectrum, we need first to make a change of variables and select   convenient linear combinations of the fermions $\chi_R$ and $\zeta_R$. The part of the action in which they appear, rescaled by the coupling $1/e^2$, is:
\bea
S_{\rm eff}
 &=&  \frac{n}{4 \pi \Lambda^2}
 \int d^2 x  \Big\{ \bar{\chi}_{R}\,i\,\pt_{L}\,  \chi_R
+\bar{\chi}_{L}\,i\, \pt_{R}\, \chi_L
+ e^2_{\chi}  \, \bar{\zeta}_L \, i
\, \pt_R \, \zeta_L  \Big. \nonumber \\ [3mm]
  && \   \Big.  -   2 \Lambda \, \bar{\chi}_{R} \Big(\chi_L - \frac{
2 \omega \pi \Lambda }{n} \zeta_L\Big) +{\rm h.c.}  \Big\}
\eea
We can diagonalize the fermion mass making a change of variable from $\chi_L$, $\zeta_L$ to $\widetilde{\chi}_L$, $\widetilde{\zeta}_L$:
\bea
\widetilde{\chi}_L  &=& \frac{1}{\sqrt{ 1+u}}\left( \chi_L -  c \zeta_L \right) \ , \nonumber \\
\widetilde{\zeta}_L  &=& \frac{1}{\sqrt{1+u}} \left(  c^{-1}\chi_L +  \zeta_L \right)  \ ,
\label{change}
\eea
where $c = 2 \omega \pi \Lambda / n$. In this new basis,
$\widetilde{\zeta}_L$ is  totally decoupled, and we can just ignore it.

Furthermore,  we want to factorize out the $n$ factors, and also write explicitly the quadratic expansion of the potential. The outcome is
\bea
S_{\rm eff}
&=& \frac{n}{4 \pi \Lambda^2}
 \int d^2 x \left\{
-\frac1{4}F^2_{kj} + 
 |\pt_k\sigma|^2
\right.
\nonumber\\[3mm]
 && + \     
  \bar{\widetilde{\chi}}_{L}\,i\,\pt_{R}\,  \widetilde{\chi}_L
+\bar{\chi}_{R}\,i\, \pt_{L}\, \chi_R - \left[  2 \Lambda \, \sqrt{ 1 + u }\,\bar{\chi}_{R}\widetilde{\chi}_L  +{\rm h.c.}\right]
 \nonumber\\[3mm]
 && 
\left. - 4 \Lambda^2 (1 + u)(\Re\,\sigma)^2 - 4 \Lambda^2 u \ ( \Im\,\sigma)^2 + 2 \sqrt{2} \Lambda\,(\Im\,\sigma)\,\widetilde{F}
  \right\} \ .
\label{effaction}
\eea
We thus get that the fermions $\chi_L$ and $\widetilde{\chi}_R$ get  together to form a massive Dirac fermion:
\beq
m_{\chi_{R}}=m_{\widetilde{\chi}_{L}}= 2\Lambda \sqrt{ 1 + u  } \,.
\label{lmass22}
\eeq
From the quadratic term of the potential $V(\sigma)$,  we calculate the mass of the real
part of the $\sigma$ field,
\beq
m_{{\rm Re} \  \sigma}=  2\Lambda \sqrt{ 1 + u} \ .
\label{smass22}
\eeq
The anomalous $\left(\Im \ \sigma \right) F^{*}$ mixing in (\ref{effaction}), and the explicit potential for $\Im\, \sigma$, give masses to both photon
and the imaginary part of $\sigma$:
\beq
m_{\gamma}=m_{{\rm Im} \ \sigma}=  2\Lambda \sqrt{ 1 +  u  } \ .
\label{phmass22}
\eeq
The way to get this result is just a simple generalization of the argument of \cite{Witten:1978bc}. 
%We need to consider the quadratic term of the Lagrangian in $A_k$ and $\Im \, \sigma$
%\beq
%\left(
%\begin{array}{cc}
%A_k & {\rm Im} \  \sigma \\ 
%&
%\end{array}
%\right)
%\,
%\left(
%\begin{array}{cc}
%-\eta_{kj}p^2 + p_k p_j & \sqrt{2} \Lambda \epsilon_{ji}  p_i \\ 
%\sqrt{2} \Lambda \epsilon_{ki}  p_i   & p^2 + 4 \Lambda^2 u
%\end{array}
%\right)
%\,
%\left(
%\begin{array}{cc}
%A_k & \\ 
%{\rm Im}\  \sigma &
%\end{array} \ ,
%\right)
%\eeq
%and then invert the matrix to find the propagator. 
The pole in the propagator is $2\Lambda \sqrt{1+u}$.

We see that all fields from the  gauge multiplets have the same mass $2 \Lambda \sqrt{ 1 +  u  } $ in accordance  with \ntwot supersymmetry.
It is curious that, although in a different way, we obtain the enhancement of supersymmetry first hypothesized in \cite{Shifman:2005st}.
One may argue that this is just a consequence of the many approximations we are using here: first of all, the large $n$ limit; second the limitations on $\mu$.
But the supersymmetry enhancement is not dependent upon these and has a more general explanation.
If we have an heterotic theory that does not break supersymmetry, and dynamically generate a mass gap, then we automatically have enhancement supersymmetry to $\N=(2,2)$. The reason is that to obtain a massive fermion we inevitably have to take a left and a right fermion. This implies an extra degeneration of multiplets. 
We thus obtain the effect of {\it dynamical supersymmetry enhancement}. 
The reason behind that is not so different from that of \cite{Shifman:2005st}. With the transformation (\ref{change}), we have effectively decoupled the internal and the translational sector. The enhancement of supersymmetry, if supersymmetry is unbroken, is then unavoidable.

Let us now discuss an issue about the quantum phase of the $\vp$ fields. 
The photon is massive, but we have confinement. How is that possible?
The issue here is hidden in the $1/n$ approximation.  At the leading order in the $1/n$ expansion, we do not see confinement because we are just expanding around the vacuum. The difference if the other discrete vacua is a subleading term, and this is why confinement, which is certainly a feature of this theory, is not visible at the leading order in the $1/n$ expansion. 
\begin{center}
 * * *
\end{center}
We now discuss the limits of validity of the present computation. 
We essentially have to satisfy the following two conditions:
\beq
\Lambda_{3+1} \ll \sqrt{\xi} \ , \qquad \mu \ll \sqrt{\xi} \ .
\eeq
The first one is clear. We want to break the gauge group at high energy, when the coupling is still weak. Then we can compute the effective action on the vortex, just relying on a classical zero modes analysis. The rest is done by the quantum dynamics of these zero modes, which are described by the $1+1$ action.

The second condition, $\mu \ll \sqrt{\xi}$, is also important.  
As we move away from this satisfy region, and $\mu$ becomes comparable or even bigger than $\sqrt{\xi}$, there are many features that make  the effective action (\ref{cpg02}) unreliable. 
First of all, the zero mode $\xi$ has a different shape, and width, with respect to the heterotic $\CP$ ones.  
That means that we should take into account   the modification of the coefficient in front of the kinetic term. 
The superpotential $W_{1+1}$ will also have corrections with respect to the four-dimensional one.
But, most importantly, it is not clear how to make sense of the $1+1$ effective actions. 
%As the with of one of these zero modes becomes larger, higher derivatives also should be considered in the game \cite{Shifman:2008wv}.
The next section will be devoted to the discussion of what happens in this region of parameters.

%% file: d.tex
The final corner is the one in which the theory is deeply $\N=1$, and the Fayet-Iliopoulos breaks the gauge group when it is still in the weak coupling regime.
This is also the most controversial, and less studied region. 
Apparently it would seem the easiest case; everything is weakly coupled. 
But there are problematic issues, on which, we hope, 
we can shed light in what follows.

Let us start with the extreme limit, $\mu \to \infty$. 
We can forget about the adjoint field; the theory is just pure $\N=1$ SQCD. 
The FI term breaks the gauge group at high energies, where the gauge coupling is still small. Thus, there are no  strong coupling ambiguities of \cite{Bolognesi:2008sw} about the vacuum. 
The moduli space is that of classical pure $\N=1$ SQCD, deformed by the FI term.
With $\mu$ very big, but finite, we would observe a small $\propto 1/\mu$ lifting of the Higgs branch.

Now we sit at the base $\tq =0$ of this manifold, and consider the non-Abelian vortex. 
The non-Abelian vortex has thickness $1/\sqrt{\xi}$, and it has both translational and orientational zero modes. 
The vortex preserves half of the supersymmetries, $\N=(0,2)$ on its worldsheet. This implies that we have also the fermionic zero modes $\zeta_R$ and $\xi_R^a$, partners, respectively, for the bosonic translational and orientational moduli. 
The other ones, $\zeta_L$ and $\xi_L^a$, become  non-normalizable in the $\mu \to \infty$ limit. 
They behave like $1/r$ at large distance, and this is directly related to the presence of the extra massless moduli \cite{Shifman:2008wv}, the Higgs branch. 
Since these modes are non-normalizable, we are not allowed to consider them in a $1+1$ effective action. 
Their kinetic term would be infinite, and thus it would be infinitely costly to excite them.

On the other hand, a $1+1$ effective theory consisting only of the $z$, $\vp^a$ and their heterotic superpartners $\zeta_R$ and $\xi_R^a$, is not consistent on its own. 
It suffers from a sigma model anomaly \cite{quantumvortexstring}. In the gauge formulation, this is simply the anomaly for the auxiliary gauge boson $A_{\mu}$.
On one hand, we are not allowed to consider the non-normalizable modes in the effective action; on the other hand, we cannot write a consistent effective action without them.
%\footnote{This issue was mentioned in a footnote of \cite{quantumvortexstring}. It is always worth to read footnotes.}
This {\it is} a puzzle. 
%That is the problem we want to solve in what follows.

There is also another puzzle, apparently unrelated to the previous one but, as we shall see, has the same origin.
The analysis of corners (A), (B), and (C)  gives a consistent picture, with basically, the following result. 
The dynamics of the heterotic vortex, strongly {\it  depends} upon the linear term in the superpotential (\ref{sup}). 
For a generic value of $a$, supersymmetry is broken, and the phase of the moduli $\vp$ is confining. 
Something special happens when the linear term is set to zero; supersymmetry is still broken but we have $n$ discrete degenerate vortices; $\vp$ excitations are massive and not confined. 
Something else happens at the special value $a \sim \Lambda e^{i 2\pi k / n} $, when supersymmetry is restored, and also enhanced. 
These results have been consistently checked independently in the three different corners (A), (B), and (C).

This dependence upon the linear term is quite embarrassing from the point of view of the region (D). 
Since the adjoint field $\Phi$ is completely decoupled in this regime, we do not expect all these different possibilities. 
The questions 1) and 2), about being or not the vortex BPS and about the quantum phase of the $\CP$ moduli,  should have only one answer, and not depend on any external parameter regarding the interaction of the adjoint field with our low-energy SQCD.

There is only one plausible answer to these issue:
{\it The dynamical scale $\Lambda_{\CP}$, in the $\mu \to \infty$ limit, must go to zero}. 
That means that there is no running of the $1+1$ coupling constant. 
$\beta$, which stays frozen to the classical value $\beta_0$, from $\sqrt{\xi}$ all the way down to the infrared. 
This is the only possible answer to the previous puzzle; all the different phases that depend  upon the linear term $a$  coalesce into a unique one in the corner (D).  
In what follows, we shall explain the main physical reason for the freezing of the coupling. 
%This will also elucidate about the  puzzle of the sigma model anomaly. 

\begin{figure}[h!t]
\epsfxsize=11cm
\centerline{\epsfbox{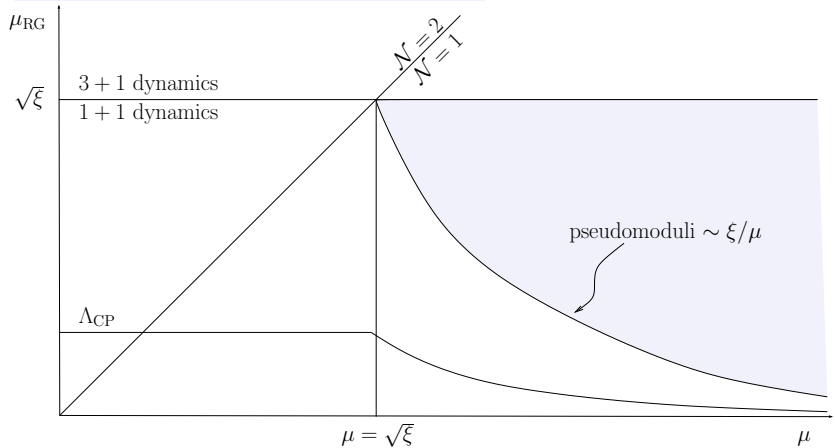}}
\caption{{\footnotesize A sketch of the RG flow (the energy scale $\murg$) as a function of the adjoint field mass $\mu$. The lines in the graph have various physical meaning described in the text.}}
\label{flow}
\end{figure}

We can elucidate better with the help of Figure \ref{flow}, where we plot various physical scales in a $\mu\ $vs.$\ \murg$ graph, where $\mu$ is the mass of the adjoint field and $\murg$ the energy scale of the RG flow.
We are in the semi-classical region $\sqrt{\xi} \gg \Lambda$. 
Physics until $\mu \sim \sqrt{\xi}$ is well described by the heterotic vortex theory, the analysis of corner (C). 
Note that the diagonal line $\murg \sim \mu$ is that where the four-dimensional theory passes from the $\N=2$ description to the $\N=1$ one. 
The vortex theory is essentially unaffected as long as $\mu \ll \sqrt{\xi}$. 
In other words, nothing special happens when the $\Lambda_{\CP}$ line crosses the diagonal line. 
For energy scales $\murg$ bigger than $\sqrt{\xi}$,  physics is four dimensional. 
At a lower scale, all the four-dimensional degrees of freedom are frozen, and the only non-trivial dynamics can happen on the $1+1$ vortex worldsheet.

Instead, the transition that  happens when the $\sqrt{\xi}$ line crosses the diagonal is very important. 
From now on, the fate of the fermionic zero mode $\xi_L$ gets separated from that of the CP moduli. 
Its width grows, and it is of order $\mu/\xi$.  
Now we have to distinguish three different regimes in the RG flow. 
Physics at energies above $\sqrt{\xi}$ is four-dimensional. 
Physics at energies below $\xi/\mu$ is $1+1$ dimensional. 
In the intermediate region, the shadowed zone in Figure \ref{flow}, we are in a hybrid situation. 
As we said, we cannot write an independent $1+1$ effective action, since the fermionic zero mode still cannot be active. And they are crucial in order to have a consistent theory.
The four dimensional gauge dynamic, on the other hand, is frozen since the gauge bosons are massive. 
But physics is, in some sense, still four dimensional. 
There are in fact massless moduli in this range of parameters. 
The dynamical scale of the $1+1$ CP moduli is frozen. 
In other words, $\beta$ is frozen to its classical value $\beta_0$, and is patiently waiting for the fermionic mode to enter in the game. 
$\beta$ starts to  run only at scales below $\xi/\mu$.

\begin{figure}[h!t]
\epsfxsize=9.5cm
\centerline{\epsfbox{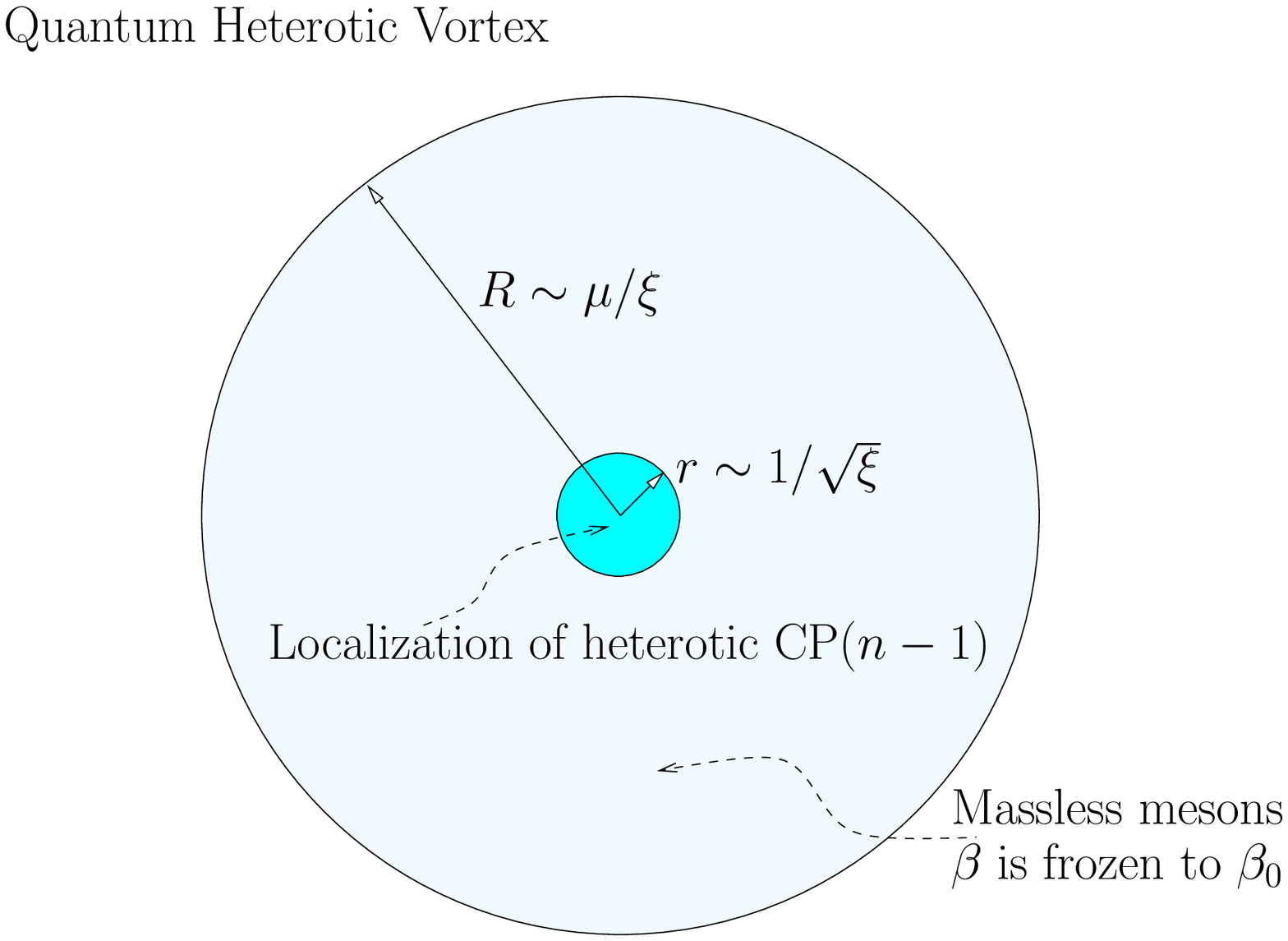}}
\caption{{\footnotesize The quantum heterotic vortex has two important scales.}}
\label{tube}
\end{figure}

In Figure \ref{tube}, we have a sketch of the two important length scales $1/\sqrt{\xi}$ and $\mu / \xi$ for the quantum heterotic vortex.
The CP heterotic zero models, $\vp^a$ and $\xi_R^a$, are localized in the core of radius $\propto 1/\sqrt{\xi}$. This vortex-core is immersed in a bath of almost-massless $3+1$ moduli. At the radius $\propto \mu/\xi$, this moduli can be integrated out. This is also the scale of the zero mode $\xi_L^a$.  
\begin{center}
 * * *
\end{center}
We now give the physical reasons  behind  this freezing effect. 
In the regime we are discussing, we have some massless CP modes confined inside the vortex, and then we have other massless modes, four dimensional, that are not confined but can, as we shall see, interact with the CP ones. 
It is clear that the vortex   cannot have an impact on the four-dimensional dynamics while the opposite is instead not true. 
The $1+1$ dynamic  is strongly influenced, as we shall see dominated, by the surrounding four-dimensional fields.

We are interested in scales of energies below $\sqrt{\xi}$ and above $\xi/\mu$. 
The gauge bosons are massive, and thus the four-dimensional dynamics is very simple, just the massless fields of the Higgs branch.
The coupling of the CP model $\beta_0$ is given by (\ref{betaatxi}). 
That means that it is a function of the four-dimensional coupling $g$ computed at the scale $\sqrt{\xi}$ set by the four-dimensional FI term. 
As long as the massless fields are there, the $\beta$ cannot run, otherwise this would imply that the dynamics of the vortex influence the bulk parameter which, as we said, are fixed in these energies scales.

We can see this effect with a more direct computation of the vortex effective theory.
This action has the following structure:
\beq
S_{\rm eff} = S_{\rm 3+1} + S_{\rm 1+1} +S_{\rm int} \ .
\label{actionmixed}
\eeq 
It is the sum of a $3+1$ action for the bulk massless moduli, plus a $1+1$ action defined on the string worldsheet that describes the orientation and translation moduli, plus an interaction term.
String moduli interact  with the surrounding massless moduli in the bulk. That means that we are not allowed to consider $S_{1+1}$ in isolation, and make the running of the coupling constant as if it were an independent $1+1$ system.
We need to take into account the interaction with the surrounding bath of $3+1$ moduli. That dominates the dynamics and keeps the $\beta$ frozen, until also the $3+1$ moduli become  massive and can be integrated out. The dynamical scale is thus 
computed starting the running only at the scale $\xi/\mu$, width the same $\beta_0$ defined by (\ref{betaatxi}). We thus have
\beq
\Lambda_{\rm CP}  \sim \Lambda_{3+1} \frac{\sqrt{\xi}}{\mu} \ .
\label{suppression}
\eeq
$\Lambda_{\rm CP}$ goes to zero in the $\mu \to \infty$ limit.

The procedure to compute $S_{3+1}$ and $S_{1+1}$ is  straightforward, and we can recall it in the following way. 
We take the expansion of the fields around some classical solution:
\bea
q(r) &=& q^{\0}(r)  + \delta_m q(r)   \nonumber \\
\tq(r) &=& \tq^{\0}(r)  + \delta_m \tq(r)  \nonumber \\
A_k(r) &=& A_{k}^{\0}(r) + \delta_m A_k(r) \ ,
\label{fluctuations}
\eea
where the ``$(0)$'' refers to the classical solution, and $\delta_m$ is a generic fluctuation of a moduli $m$, and $\delta_m q(r)$, $\delta_m \tq(r)$, $ \delta_m A_k(r)$ are the respective zero modes. 
For the case of $S_{3+1}$, we expand around the vacuum, and the moduli $m$ are the residual massless fields of the Higgs branch. 
For the case of $S_{1+1}$, we expand around the vortex solution, and the moduli are the orientational and translational ones.
We then insert this into the original action, and expand in powers of $\delta m$. If $m$ is an exact moduli, there will be no explicit dependence on $m$ but only on its derivatives. 
The coefficient in front of the kinetic term $\partial m   \partial m$ is then computed as function of the respective zero modes.

To obtain $S_{\rm int }$ is instead not straightforward; we have to deal with some renormalization issues. 
We have to expand around the vortex solution, but we have to consider simultaneously the fluctuation of the vortex moduli $\varphi$, $z$, and of that of the bulk massless fields $\delta \tq$. 
But as soon as we excite the $\tq$ field, a logarithmic tail comes out of the vortex, and makes the tension infinite.
That is certainly a signal that the vortex, and presumably also its moduli, interacts with the surrounding massless fields. But to make a sensible computation, we need some regularization procedure.

To obtain a finite result, we need to consider a fluctuation of $\tq$ with some wavelength different from infinity. 
The tension of the vortex, and consequently also the interaction terms $S_{\rm 1+1}$, will depend on this scale. 
In particular, they all diverge if we try to change $\tq$ homogeneously in all the space, at zero wavelength.
To simplify at most the computation, we can just consider the space as if it is cylindrical and compact, with a cutoff radius $1/\lambda_{\perp}$. The physical interpretation is that $\lambda_{\perp}$ is the transverse energy scale of the $\tq$ fluctuation.
In this way, we have regularized the zero modes $\delta q$, $\delta \tq$, and $ \delta A_k$ corresponding to the massless bulk fluctuation. Otherwise, they all would have a $\log$ divergence in the IR far from the vortex core.

We need to make an important distinction between parallel and orthogonal excitations. Since we are dealing with a non-Abelian vortex, we have to choose some particular orientation for the $\vp$ field to put in the zero solution of (\ref{fluctuations}). The massless bulk fluctuation $\delta \tq$ has also an orientation, and can be parallel or orthogonal to the $\vp$ field. 
The shape of the zero modes strongly depends upon this choice, and this, ultimately, is the cause of the interaction between $\vp$ and $\delta \tq$.

\begin{figure}[h!t]
\epsfxsize=8cm
\centerline{\epsfbox{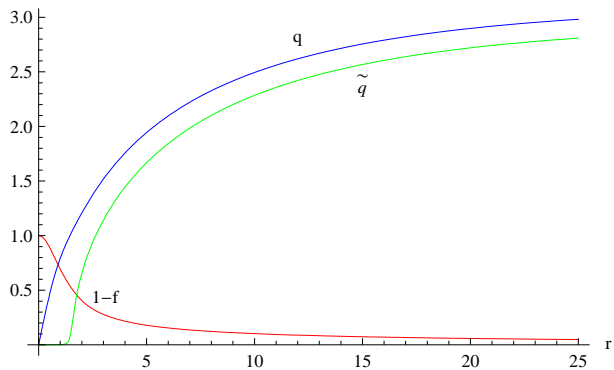}}
\caption{{\footnotesize Vortex with $\delta \tq$ excitation {\it parallel} to its orientation. }}
\label{paral}
\end{figure}

After the choice of parallel orientation (\ref{vortexconf}), the problem can be reduced to the following Abelian model:
\beq
{\cal L} = -\frac{1}{4e^2}F_{\mu\nu}F^{\mu\nu} + |(\partial_\mu - i A_\mu) q|^2 + |(\partial_\mu + i  A_\mu) \tq|^2   - \frac{e^2}{2}( |q|^2 - |\tq|^2  -2\xi)^2  \ , \ . 
\label{BPSabelian}
\eeq
where $\alpha =-1$ in the parallel case and $\alpha=0$ in the orthogonal case. 
The ansatz for the profile functions is
\beq
A_k=-\epsilon_{kl}\frac{\hat{r}_l}{r}f(r) \ , \qquad q = e^{i\theta}\sqrt{2{\xi}} \, q(r) \ , \qquad \tq = e^{- i \theta}\sqrt{2{\xi}} \, \tq(r) \ .
\eeq
The solution of Figure \ref{paral} is obtained with this model for the parallel case $\alpha=-1$, and with the choice of parameters $e=1$, $\xi=.5$.

In the parallel case, we have the log tail typical of vortices with flat directions.
If $\lambda_{\perp} \ll \sqrt{\xi}$, we can approximate them outside the core of the vortex with
\beq
\delta q(r) \sim  \delta \tq^{\dagger}\log \left( r \lambda_{\perp}\right)  \ , \qquad 
\delta \tq(r) \sim  \delta \tq \log \left( r \lambda_{\perp}\right) \ , 
\qquad  
\delta A_k(r) \sim 0 \ ,
\eeq
where we used $\delta \tq$ as our $\delta m$. 
What happens inside the core is not very important, as long as we consider a small enough cutoff $\lambda_{\perp}$. 
The only important thing to know is that the vortex core regularize the log divergence in the core, and all the fields go to zero. We show an example  in Figure \ref{paral}.

We can now take the fluctuations (\ref{fluctuations}), with all the zero models $z$, $\vp$, and $\delta \tq$ in the vortex background, and insert this back into the Lagrangian (\ref{lagvortice}).
The term we are interested in comes out very easily, just from the term:
\bea
S &=&\int d^2x \  T_{\rm V}  + \dots  \nonumber \\ &=& \int d^2 x \left(4 \pi \xi + \log \left( \frac{\sqrt{\xi}}{\lambda_{\perp}} \right) |\delta \tq|^2 \right) + \dots \ .
\label{inte}
\eea
This is the tension of the vortex, integrated over its worldsheet. 
In general, it is a constant term  that does not interfere with the higher-order terms containing the moduli interaction. In the case at hand, the log tail of the scalar fields modifies the vortex tension, and this introduces a mass term of the moduli $\delta \tq$.

Things are very different if  $\delta q$  is instead orthogonal to the vortex orientation.
We can see from (\ref{BPSabelian}) that the vortex sector $A_{mu}$, $q$ and the flat direction
are completely decoupled. It is so indifferent if we make the fluctuation $\delta \tq$ around the plain vacuum or around the vortex: 
\beq
\delta q(r) \sim 0  \ , \qquad 
\delta \tq(r) \sim \delta \tq \ , 
\qquad  
\delta A_k(r) \sim 0 \ .
\eeq

%\begin{figure}[h!t]
%\epsfxsize=8cm
%\centerline{\epsfbox{orto.eps}}
%\caption{{\footnotesize Vortex with $\delta \tq$ excitation {\it perpendicular} to its orientation. }}
%\label{orto}
%\end{figure}

We can thus extract from (\ref{inte}) the interaction term between the orientational moduli $\vp$ of the non-Abelian string  and the bulk fluctuation $\delta \tq$ of the Higgs branch:
\beq
S_{\rm int} \sim    \log \left( \frac{\sqrt{\xi}}{\lambda_{\perp}}  \right) \frac{1}{\beta}\int d^2 x \   \left| \vp^l \ \delta  \tq_l \right|^2  \ .
\label{interazione}
\eeq
Note that if we consider the renormalization of the coupling constant $\beta=\beta_0 \log(\murg/\sqrt{\xi})$, and make the renormalization scales (transversal $\lambda_{\perp}$ and longitudinal $\murg$) equal, we get
\beq
S_{\rm int} \sim    \frac{1}{\beta_0}\int d^2 x \   \left| \vp^l \ \delta  \tq_l \right|^2  \ .
\label{interazionefinale}
\eeq
\begin{center}
 * * *
\end{center}
We now give our argument for the existence of the abovementioned conformal window.
In the window of energy scales below $\sqrt{\xi}$ and above $\xi/\mu$, the action of the vortex is of type (\ref{actionmixed}), with a the  interaction (\ref{interazionefinale}). 
We want to write an effective action, effective from two points of view. 
One because we want to integrate out the $3+1$ degrees of freedom of the bulk Higgs branch; the second because we want to consider it at some energy scale $\murg$. The RG flow must be done scaling simultaneously the longitudinal and traversal scales, so that $\lambda_{\perp}$ in (\ref{interazione}) is essentially $\murg$. 
The effective action, constrained by the $\N=(0,2)$ supersymmetry, must be of the following kind
\bea
S_{\rm eff} &=& -\frac{i \mathcal{Z}(\murg)}{2}  \int d^2x\, d^2 \theta_R
  \ \bar\Phi_l (\partial_L- i U) \Phi^l \nonumber \\ & & + \frac{\tau(\murg)}{4} \int d^2x \, d\theta_R \ \left.\Upsilon\right|_{\bar{\theta}_R=0}  + {\rm h.c.} \ ,
\nonumber \\
&& + \frac{1}{8 e_{\Upsilon}^2} \int d^2x \, d^2  \theta_R \  \Upsilon^\dagger \Upsilon \ ,
\eea
where $ \mathcal{Z}(\murg)$ is a wave-function renormalization, obtained integrating out the $3+1$ fields. The renormalization of $\tau(\murg)$ is instead due to the $1+1$ loops, and is given by the usual logarithmic scaling.\footnote{$\Phi^l$ is the chiral superfield containing $\vp^l$ and $\xi_{R}^l$. $U$ is the vector superfield containing $D$, $\chi_L$ and $A_{\mu}$. $\Upsilon$ is a Fermi multiplet, and is the field strength of $U$. The reader can consult \cite{Edalati:2007vk} for the conventions.}
We can separate the two quantum effects because the Higgs branch fields do not communicate directly with the $\Upsilon$ fields.
The $1+1$-loop  has to cancel out with the $3+1$ quantum effects, otherwise the action is not consistent.
If $  \mathcal{Z}(\murg) \sim \tau(\murg)$, then the action is conformal and there is no dynamical scale generated. The kinetic term for $\Upsilon$ is infinite, since it is proportional to 
$1/\Lambda$, and so there are no anomaly inconsistencies.

\begin{figure}[h!t]
\epsfxsize=10cm
\centerline{\epsfbox{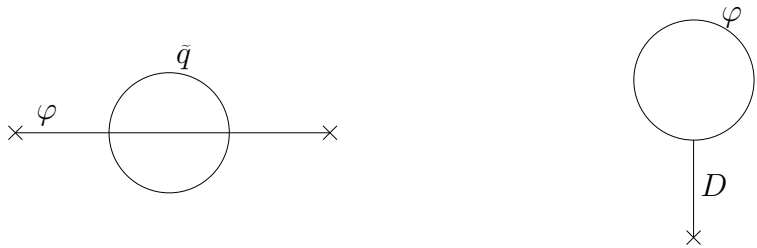}}
\caption{{\footnotesize  Two diagrams coming from the quantization of the action \ref{provaaction}. The first is the wave-function renormalization for $\vp$. It is a two-loop diagram: one loop is four-dimensional, and the other two-dimensional. The second graph the one-loop renormalization of the coupling $\beta$; it is a tadpole for the auxiliary field $D$. }}
\label{diagrams}
\end{figure}

To see the physical reason for this cancellation, we consider the simplified action
\beq
\label{provaaction}
S  = \int d^4 x \ |\partial_{\mu} \tq_l|^2  + \int d^2 x \left\{ |\nabla_{k} \vp^{l}|^2  + D (|\vp^{l}|^2-\beta) + \frac{1}{\beta} \left| \vp^l \,   \tq_l  \right|^2 \right\}
 \ ,
\eeq
in which we have just taken into account the scalar fields $\vp$ and $\tq$. We have also simplified because we do not consider, in full detail, the Higgs branch and its metric. But for the effect we are interested in, this toy-model captures all the essential features.
The renormalization of the wave-function  of the $\vp$ is given by the first diagram in Figure \ref{diagrams}. It is a $2$-loop diagram involving two interaction vertices with the four-dimensional field $\tq$, and three propagators. 
One of the loops is four-dimensional; the other is two-dimensional. In total, the divergence is a logarithmic one $\int d^6k \frac{1}{k^6} \sim \log \murg$.
The renormalization of the coupling constant is given by the second diagram in the Figure. It is a tadpole for the auxiliary field, and it involves one two-dimensional loop and one propagator. It is logarithmic divergent $\int d^2k \frac{1}{k^2} \sim  \log \murg$. And gives the usual logarithmic running of the coupling $\beta$. The {\it physical} coupling is obtained by $\beta$ divided by the wave-function renormalization of $\vp$, and thus the theory remains conformal. This is the same cancellation that happened from (\ref{interazione}) to (\ref{interazionefinale}).
The kinetic term for the auxiliary field $A_{\mu}$, involves one two-dimensional loop of the field $\vp$, with two propagators. It has no UV-finite, but IR-divergent: $\int d^2k \frac{1}{k^4}$. If the theory generate a mass gap, then we usually put $\Lambda$ as the IR-cutoff, and thus we get the quantum kinetic term for $A_{\mu}$ (\ref{e22}).
In the present situation, there is no mass generation, and the diagram is $\infty$, and that means that the gauge field is decoupled ($e_{A_{\mu}}=0$), and there are no problems for the anomaly.

%Note that in the $\mu \to \infty$ limit the physics of the quantum heterotic vortex %is at the end very simple: is not quantum at all!

%\begin{center}
% * * *
%\end{center}

%\texttt{Take the action and make the limit. Describe the issues %about the kinetic terms and the superptential.}

%% file: concl.tex
In this paper, we have considered various aspects of the quantum heterotic vortex, for number of colors equal to the number of flavors $n_c =n_f =n$. 
We focused in particular on four regions  in the parameter  space  spanned by the mass $\mu$ and the FI term $\xi$ (Figure \ref{box}). 
These are the four regimes where quantum effects can be, somehow, treated using known techniques. 
We found agreement between the various approaches, in particular with respect to the qualitative feature about being or not the string BPS, and about the quantum phase of the internal $\CP$ moduli.

Clearly, the FI parameter, and the mass for the adjoint field, have a big impact on the dynamics of the theory. 
Physics changes considerably, as we move in the parameter space spanned by $\mu$ and $\xi$. 
Many qualitative feature, though, remain the same and can be checked, and compared using the various approximations.
A physical idea that permeates the paper  is that strings that we detect in the four-dimensional approaches, such as (A) and (B), are nothing by the remnants, or ground-states, of the quantum dynamics of the non-Abelian string that we find in the worldsheet approaches, such as (C) and (D).

The goal, for the moment, was to provide an overall picture of  the entire parameter space.
There are  aspects of this paper, though, that deserve a more detailed discussion. 
Two are of particular importance.

The analysis of corner (B) in Section \ref{b}  provides the mechanism to understand the $n$-fold degeneracy from the point of view of the meson-baryon description. 
Essential is the quantum deformation of the moduli space (\ref{qdef}) that triggers this degeneracy.  
Although the argument is solid, we still lack   an elegant way to write an effective theory that incorporates both the $\U(1)_B$ gauge, the baryons and mesons, and the quantum deformation (\ref{qdef}). The example (\ref{superpottoy}) is valid in the extreme limit of $n=1$. Although it captures some  essential effects, like the stability of the log tail and the meson condensate in the interior of the vortex, it lacks incorporation of the degeneracy (\ref{degeneracy}) in a dynamic  fashion. 
A proper treatment for $n>1$ is  highly desirable.

The analysis of corner (D) is not yet complete. 
We were able to show that in this regime the vortex zero modes {\it interact} with the external bath of $3+1$ particles. 
This implies that the $1+1$ effective theory should not be considered in isolation, and in particular that the $\beta$ coupling should not run just according to the $1+1$ loops computations. 
We conjectured that $\beta$ remains frozen to $\beta_0$, from the scale $\sqrt{\xi}$ of the Higgs breaking, down to the scale $\xi/\mu$ of the light mesonic fields. 
%These facts should solve some puzzles we discussed. In particular that fact that the vortex properties should depend on the linear term $a$ in the superpotential, despite the absence of the $\Phi$ field in this regime. Another puzzle is that the $1+1$ effective theory, without the both left and right fermions, is not quantum mechanical consistent, if treated in isolation. And only at scales below $m_{\rm l}$ we are allowed to consider both left and right fermions in the $1+1$ theory.

The freezing of the coupling constant is something that we need for consistency, and we also gave  physical arguments for that. 
%A rigorous proof for that effect,  would require an explicit an explicit quantum renormalization of the action (\ref{actionmixed}). 
%, with a mixture of $3+1$ and $1+1$ degrees of freedom interacting with each other. 
%Let us state here clearly what are the proved facts, and what are the conjectures. 
%It is a fact that the action (\ref{actionmixed}) has some puzzling aspects.
First of all, the $S_{1+1}$ theory by its own is not a consistent QFT, since we have only left fermions and no right ones.  There is thus an anomaly for the gauge field $A_{\mu}$.
Even if we forget about the other terms in (\ref{actionmixed}), we cannot quantize $S_{1+1}$ in isolation.
Another fact is that the answers to the two questions posed in the Introduction  depend upon the linear term in the superpotential. And this is hardly understandable from the point of view of the deep corner (D), pure SQCD without the adjoint field. 
The existence of the interactions $S_{\rm int}$
between the $1+1$ and $3+1$ degrees of freedom  is the crucial point in order to solve these puzzles.
This is what should make the theory consistent and anomaly free.
The {\it conjecture} is that the only way to make a consistent, gauge-anomaly-free, quantization of (\ref{actionmixed}), we need to have a conformal window, between the scales $\sqrt{\xi}$ and $\xi/\mu$ where the coupling $g$ does not run. 
We gave some heuristic arguments for why, and how, this cancellation happens.

%Second, even if we ignore the previous problem and run the coupling $\beta$ as if it was an independent $S_{1+1}$ action, then we would find a mass gap generation in the low-energy. The last is clearly not possible.  We showed that the $\vp$ field interact with the $\delta \tq$. That menas that the sigma model on the $1+1$ worldsheet interacts with the sigma model in $3+1$, the theory of the Higgs branch. Since we know that te sigma model in $3+1$ has no mass generation (the Higgs branch is at most modifyed in its metric structure, but remains a flat direction), that imply that also the $\vp$ fluctiation cannot developp a mass gap.

%\texttt{3+1 loops are dominant}

There also other aspects of corner (D) that deserve  more careful analysis. The interaction (\ref{interazione}) should also be formulated in a supersymmetric way. A possibility could be the following $\N=(0,2)$ superpotential-like term
\beq 
S_{\rm int} \sim    g_0
 \int d^2 x d \theta_R    \   \left.\Psi_{q}^i \   \Phi_l \,  \tQ^{\,l}_i  \right|_{\bar{\theta}_R=0} + {\rm h.c.} \ ,
\label{interazionesusy}
\eeq 
where $\Psi^i$ is the Fermi multiplet containing $\psi_{q \, L}$, $\Phi_l$ the chiral superfield containing $\varphi_l$, $\xi_{l \, R}$ and $\tQ$ the chiral superfield containing $\tq$, $\psi_{\tq \, L}$. 
It should be possible also to write it in a geometric way, as an interaction between the geometric $\CP(n-1)$ manifold on the worldsheet  and the moduli space of the Higgs branch in the bulk.
It is also true that this is not the unique interaction. Other interactions should be present, in particular, the ones that involve  the translational sector.

The next step in the heterotic vortex saga  is that of $n_f$ higher than $n_c$. In trying to generalize these ideas to higher $n_f$, we encounter one major obstacle, already at the $\N=2$ level (i.e., without $\mu$ breaking terms). 
There are extra moduli, and a Higgs branch, already at $\N=2$. 
The non-Abelian vortex becomes semi-local, and the radius is no longer fixed at $1/\sqrt{\xi}$. It is not clear how to make sense of an effective vortex theory in these circumstances.
It is not clear how to disentangle the $3+1$ dynamics from the $1+1$.
The situation we encountered in this paper in corner (D), may be instructive in this respect. 
The massless moduli have the effect of freezing the dynamics of the $1+1$ vortex. 
But there is also an important difference, which we never encouter in the present paper:  semilocal vortices have size as a bosonic zero mode. 
%A thing we can anticipate is that the superoptential (\ref{supsenzalin}), without the linear term, does not break the supersymmetries of the vortex. That is quite clear from corner (A) and (B).
%But there is a lesson that should go trough. Non-normalizable zero modes are the signal of some massless residual $3+1$ degree of freedom. 
%The proper way to deal with them, is to write a mixed action like \ref{actionmixed}, where they are hidden inside the $3+1$ part.

We can anticipate a few of the new features of the $n_f > n_c$ case.
The equivalent analysis for corner (A), the perturbed $\N=2$, is a straightforward generalization of what we have done in Section \ref{a}. 
The SW curve, at the root of the baryonic branch, factorizes as:
\beq
{\cal P}_{(n_c,n_f)}(z)  =    \frac{1}{4} z^{2(n_f - n_c)} \left(z^{2n_c - n_f} - \Lambda^{2n_c - n_f} \right)^2 \ .
\eeq 
The difference with respect to (\ref{factorizationbb}) is that there are $2(n_f - n_c)$ roots located at $z=0$. The others are located in a $\Z_{2n_c -n_f}$ symmetric way at $z=\omega_{2n_c -n_f} \Lambda$. 
The roots located at zero gives a low-energy $\SU(\tnc)$ theory, where $\tnc= n_f - n_c$, with $n_f$ fundamentals hypermultiplets;
it is infrared free, since $2n_f > \tnc$ \cite{Argyres:1996eh}.
With the superpotential (\ref{supsenzalin}) and the FI term, we have a BPS vortex 
The other Abelian vortices, coming from the $\Z_{2n_c -n_f}$ symmetric roots, are instead non-BPS and behave  much in the same way the ones we discussed in this paper.

The relation between corner (A) and corner (C), for example, should thus provides an insight into the dynamics of semilocal vortices. 
Their moduli space has been widely studied from a semi-classical perspective \cite{Hanany:2004ea,Shifman:2006kd,Eto:2007yv}.
In corner (C) we have a semilocal vortex with $(n_c,n_f)$. In corner (A) we have a BPS semilocal vortex with $(\tnc,n_f)$, plus $2n_c - n_f$ abelian non-BPS vortices. What we find in the corner (A) analysis, should correspond to the ground states of the theory in corner (C).  It looks as if there is a $\CP(2n_c-n_f)$ sub-sector of the semilocal vortex which goes into strong coupling effect, plus another sector which contains all the semilocal moduli, which has no strong coupling effect and shows in the low-energy as a semilocal BPS $(\tnc,n_f)$ vortex. 
The explanation for this, could reside in the $\tq$ flat directions, much in the same way of corner (D) in Section \ref{d}.

%The Higgs branch is a complex manifold with dimension $2(n_f - n_c)$. The base of the manifold, $\tq = 0$, is a compact $\CP(n_f - n_c)$ space, and is the one responsible for the semi-local vortices. We have also other $n_f - n_c$ non-compact flat-directions, due to the $\tq$ fields. 

%In particular they becomes BPS if we choose a linear term in the superpotential (\ref{sup}) with $a \sim \Lambda e^{(i \, 2\pi k)/(2n_c-n_f)}$.  

%Is still not clear how to interpret these vortices in the descriptions of corners (B) and (C).

%% file: nota-heterotic-new.bbl
\begin{thebibliography}{99}
{\small \itemsep -2pt }

 \bibitem{Hanany:2003hp}
  A.~Hanany and D.~Tong,
  ``Vortices, instantons and branes,''
  JHEP {\bf 0307}, 037 (2003)
  [arXiv:hep-th/0306150].

\bibitem{Auzzi:2003fs}
  R.~Auzzi, S.~Bolognesi, J.~Evslin, K.~Konishi and A.~Yung,
  ``Nonabelian superconductors: Vortices and confinement in N = 2 SQCD,''
  Nucl.\ Phys.\  B {\bf 673}, 187 (2003)
  [arXiv:hep-th/0307287].


\bibitem{futuro} S.~Bolognesi, {\it to appear}


\bibitem{Bolognesi:2008sw}
  S.~Bolognesi,
  ``A Coincidence Problem: How to Flow from N=2 SQCD to N=1 SQCD,''
  arXiv:0807.2456 [hep-th].
  %%CITATION = ARXIV:0807.2456;%%





%\cite{Argyres:1996eh}
\bibitem{Argyres:1996eh}
  P.~C.~Argyres, M.~R.~Plesser and N.~Seiberg,
  ``The Moduli Space of N=2 SUSY {QCD} and Duality in N=1 SUSY {QCD},''
  Nucl.\ Phys.\  B {\bf 471}, 159 (1996)
  [arXiv:hep-th/9603042].
   

 
  
  
\bibitem{Shifman:2005st}
  M.~Shifman and A.~Yung,
  ``Non-abelian flux tubes in SQCD: Supersizing world-sheet supersymmetry,''
  Phys.\ Rev.\  D {\bf 72}, 085017 (2005)
  [arXiv:hep-th/0501211].
   

\bibitem{Edalati:2007vk}
  M.~Edalati and D.~Tong,
  ``Heterotic vortex strings,''
  JHEP {\bf 0705}, 005 (2007)
  [arXiv:hep-th/0703045].
  


\bibitem{quantumvortexstring}
   D.~Tong,
  ``The quantum dynamics of heterotic vortex strings,''
  JHEP {\bf 0709}, 022 (2007)
  [arXiv:hep-th/0703235].



\bibitem{Tong:2008qd}
  D.~Tong,
  ``Quantum Vortex Strings: A Review,''
  arXiv:0809.5060 [hep-th].



\bibitem{Shifman:2008wv}
  M.~Shifman and A.~Yung,
  ``Heterotic Flux Tubes in N=2 SQCD with N=1 Preserving Deformations,''
  Phys.\ Rev.\  D {\bf 77}, 125016 (2008)
  [arXiv:0803.0158 [hep-th]].



\bibitem{Shifman:2008kj}
  M.~Shifman and A.~Yung,
  ``Large-N Solution of the Heterotic N=(0,2) Two-Dimensional CP(N-1) Model,''
  Phys.\ Rev.\  D {\bf 77}, 125017 (2008)
  [arXiv:0803.0698 [hep-th]].
  %%CITATION = PHRVA,D77,125017;%%

\bibitem{Bolokhov:2009sg}
  P.~A.~Bolokhov, M.~Shifman and A.~Yung,
  ``Description of the Heterotic String Solutions in U(N) SQCD,''
  arXiv:0901.4603 [hep-th].
  %%CITATION = ARXIV:0901.4603;%%


%\bibitem{Shifman:2007kd}
%  M.~Shifman and A.~Yung,
%  ``Confinement in N=1 SQCD: One Step Beyond Seiberg's Duality,''
%  Phys.\ Rev.\  D {\bf 76}, 045005 (2007)
%  [arXiv:0705.3811 [hep-th]].



  
\bibitem{Auzzi:2004yg}
  R.~Auzzi, S.~Bolognesi and J.~Evslin,
  ``Monopoles can be confined by 0, 1 or 2 vortices,''
  JHEP {\bf 0502}, 046 (2005)
  [arXiv:hep-th/0411074].


   \bibitem{Bolognesi:2004yh}
  S.~Bolognesi,
  ``The holomorphic tension of vortices,''
  JHEP {\bf 0501}, 044 (2005)
  [arXiv:hep-th/0411075];
%``The holomorphic tension of nonabelian vortices,''
%  Nucl.\ Phys.\  B {\bf 719}, 67 (2005)
%  [arXiv:hep-th/0412241].

  %%CITATION = JHEPA,0501,044;%%

  
  \bibitem{Intriligator:1995au}
  K.~A.~Intriligator and N.~Seiberg,
  ``Lectures on supersymmetric gauge theories and electric-magnetic  duality,''
  Nucl.\ Phys.\ Proc.\ Suppl.\  {\bf 45BC}, 1 (1996)
  [arXiv:hep-th/9509066].
  


\bibitem{Witten:1978bc}
  E.~Witten,
  ``Instantons, The Quark Model, And The 1/N Expansion,''
  Nucl.\ Phys.\  B {\bf 149}, 285 (1979).
  %%CITATION = NUPHA,B149,285;%%

\bibitem{D'Adda:1978uc}
  A.~D'Adda, M.~Luscher and P.~Di Vecchia,
  ``A 1/N Expandable Series Of Nonlinear Sigma Models With Instantons,''
  Nucl.\ Phys.\  B {\bf 146}, 63 (1978).
  %%CITATION = NUPHA,B146,63;%%






%\bibitem{Shifman:2007ce}
%  M.~Shifman and A.~Yung,
%  ``Supersymmetric Solitons and How They Help Us Understand %Non-Abelian %  Gauge Theories,''
%  Rev.\ Mod.\ Phys.\  {\bf 79}, 1139 (2007)
%  [arXiv:hep-th/0703267].






\bibitem{Dorey:1998yh}
  N.~Dorey,
  ``The BPS spectra of two-dimensional supersymmetric gauge theories with
  twisted mass terms,''
  JHEP {\bf 9811}, 005 (1998)
  [arXiv:hep-th/9806056].


\bibitem{Dorey:1999zk}
  N.~Dorey, T.~J.~Hollowood and D.~Tong,
  ``The BPS spectra of gauge theories in two and four dimensions,''
  JHEP {\bf 9905}, 006 (1999)
  [arXiv:hep-th/9902134].

\bibitem{Shifman:2004dr}
  M.~Shifman and A.~Yung,
  ``Non-Abelian string junctions as confined monopoles,''
  Phys.\ Rev.\  D {\bf 70}, 045004 (2004)
  [arXiv:hep-th/0403149].
  %%CITATION = PHRVA,D70,045004;%%



%\cite{Penin:1996si}
\bibitem{Penin:1996si}
  A.~A.~Penin, V.~A.~Rubakov, P.~G.~Tinyakov and S.~V.~Troitsky,
  ``What becomes of vortices in theories with flat directions,''
  Phys.\ Lett.\  B {\bf 389}, 13 (1996)
  [arXiv:hep-ph/9609257].
  %%CITATION = PHLTA,B389,13;%%




%\cite{Achucarro:2001ii}
\bibitem{Achucarro:2001ii}
  A.~Achucarro, A.~C.~Davis, M.~Pickles and J.~Urrestilla,
  ``Vortices in theories with flat directions,''
  Phys.\ Rev.\  D {\bf 66}, 105013 (2002)
  [arXiv:hep-th/0109097].
  %%CITATION = PHRVA,D66,105013;%%



\bibitem{Yung:1999du}
  A.~Yung,
  ``Vortices on the Higgs branch of the Seiberg-Witten theory,''
  Nucl.\ Phys.\  B {\bf 562}, 191 (1999)
  [arXiv:hep-th/9906243].



%\bibitem{Auzzi:2007wj}
 % R.~Auzzi, M.~Eto and W.~Vinci,
 % ``Static Interactions of non-Abelian Vortices,''
 % JHEP {\bf 0802}, 100 (2008)
 % [arXiv:0711.0116 [hep-th]].
  %%CITATION = JHEPA,0802,100;%%



%\bibitem{Davis:1997bs}
%  S.~C.~Davis, A.~C.~Davis and M.~Trodden,
%  ``N = 1 supersymmetric cosmic strings,''
%  Phys.\ Lett.\  B {\bf 405}, 257 (1997)
%  [arXiv:hep-ph/9702360].


\bibitem{Ritz:2004mp}
  A.~Ritz, M.~Shifman and A.~Vainshtein,
  ``Enhanced worldvolume supersymmetry and intersecting domain walls in N = 1 SQCD,''
  Phys.\ Rev.\  D {\bf 70}, 095003 (2004)
  [arXiv:hep-th/0405175].


\bibitem{Hanany:2004ea}
  A.~Hanany and D.~Tong,
  ``Vortex strings and four-dimensional gauge dynamics,''
  JHEP {\bf 0404}, 066 (2004)
  [arXiv:hep-th/0403158].


\bibitem{Shifman:2006kd}
  M.~Shifman and A.~Yung,
   ``Non-Abelian semilocal strings in N = 2 supersymmetric QCD,''
  Phys.\ Rev.\  D {\bf 73}, 125012 (2006)
  [arXiv:hep-th/0603134].


\bibitem{Eto:2007yv}
  M.~Eto {\it et al.},
  ``On the moduli space of semilocal strings and lumps,''
  Phys.\ Rev.\  D {\bf 76}, 105002 (2007)
  [arXiv:0704.2218 [hep-th]].
  %%CITATION = PHRVA,D76,105002;%%

\bibitem{Witten:1993yc}
  E.~Witten,
  ``Phases of N = 2 theories in two dimensions,''
  Nucl.\ Phys.\  B {\bf 403}, 159 (1993)
  [arXiv:hep-th/9301042].
  %%CITATION = NUPHA,B403,159;%%


\bibitem{Tong:2005un}
  D.~Tong,
  ``TASI lectures on solitons,''
  arXiv:hep-th/0509216.
  %%CITATION = HEP-TH/0509216;%%



\bibitem{Shifman:2007ce}
  M.~Shifman and A.~Yung,
 ``Supersymmetric Solitons and How They Help Us Understand Non-Abelian   Gauge
  Theories,''
  Rev.\ Mod.\ Phys.\  {\bf 79}, 1139 (2007)
  [arXiv:hep-th/0703267].
  %%CITATION = RMPHA,79,1139;%%
  

\end{thebibliography}
